\documentclass[sigconf]{acmart}

\AtBeginDocument{%
  }

\setcopyright{acmlicensed}
\copyrightyear{2024}
\acmYear{2024}
\setcopyright{acmlicensed}
\acmConference[CCS '24] {Proceedings of the 2024 ACM SIGSAC Conference on Computer and Communications Security}{October 14--18, 2024}{Salt Lake City, UT, USA.}
\acmBooktitle{Proceedings of the 2024 ACM SIGSAC Conference on Computer and Communications Security (CCS '24), October 14--18, 2024}
\acmISBN{979-8-4007-0636-3/24/10}
\acmDOI{10.1145/XXXXXX.XXXXXX}

\copyrightyear{2024}
\acmYear{2024}
\setcopyright{rightsretained}
\acmConference[CCS '24] {Proceedings of the 2024 ACM SIGSAC Conference on Computer and Communications Security}{October 14--18, 2024}{Salt Lake City, UT, USA.}
\acmBooktitle{Proceedings of the 2024 ACM SIGSAC Conference on Computer and Communications Security (CCS '24), October 14--18, 2024}
\acmDOI{10.1145/3658644.3690257}
\acmISBN{979-8-4007-0636-3/24/10}

\usepackage{etoolbox} 

\usepackage{enumitem, array, bm, multirow}
\usepackage{amsmath,amsfonts,amsthm}
\usepackage{algorithm, algorithmic, float}
\usepackage{graphicx}
\usepackage{caption}
\usepackage{subcaption}
\usepackage{lipsum} 
\usepackage{diagbox}
\usepackage{siunitx}
\usepackage{makecell}

\newtheorem{theorem}{Theorem}[section] 
\newtheorem{definition}[theorem]{Definition}

\newcommand{\INPUT}{\item[\textbf{Input:}]}
\newcommand{\OUTPUT}{\item[\textbf{Output:}]}

\newcommand{\PRECOMPUTE}[1]{\item[\textbf{Precompute:}] #1}

\usepackage{hyperref,xcolor}
\hypersetup{
  colorlinks,
  linkcolor={green!80!black},
  citecolor={red!70!black},
  urlcolor={blue!70!black}
}

\usepackage{booktabs,xspace,url}

\newcommand{\mypara}[1]{\vspace*{0.06in}\noindent\textbf{#1} \xspace}
\renewcommand{\Pr}[1]{\ensuremath{\mathsf{Pr}\left[#1\right]}\xspace}

\newcommand{\Data}{D}

\newcommand{\Lap}[1]{\ensuremath{\mathsf{Lap}\left(#1\right)}\xspace}
\newcommand{\Gau}[1]{\ensuremath{\mathsf{N}\left(#1\right)}\xspace}

\newcommand{\dLap}[1]{\ensuremath{\mathsf{Lap}_{\mathbb{Z}}\left(#1\right)}\xspace}
\newcommand{\dGau}[1]{\ensuremath{\mathsf{N}_{\mathbb{Z}}\left(#1\right)}\xspace}
\newcommand{\tdLap}[1]{\ensuremath{\mathsf{Lap}_{\mathbb{Z},N}\left(#1\right)}\xspace}
\newcommand{\tdGau}[1]{\ensuremath{\mathsf{N}_{\mathbb{Z},N}\left(#1\right)}\xspace}

\newcommand{\proto}[1]{\texttt{#1}}

\begin{document}

\title{Benchmarking Secure Sampling Protocols for Differential Privacy}


\author{Yucheng Fu}
\affiliation{%
  \institution{University of Virginia}
  \city{}
  \country{}
  }
\email{zdp8uu@virginia.edu}

\author{Tianhao Wang}
\affiliation{%
  \institution{University of Virginia}
  \city{}
  \country{}
  }
\email{tianhao@virginia.edu}

\renewcommand{\shortauthors}{Yucheng Fu \& Tianhao Wang}

\begin{abstract}
Differential privacy (DP) is widely employed to provide privacy protection for individuals by limiting information leakage from the aggregated data. 
Two well-known models of DP are the \textit{central model} and the \textit{local model}. The former requires a trustworthy server for data aggregation, while the latter requires individuals to add noise, significantly decreasing the utility of aggregated results. 
Recently, many studies have proposed to achieve DP with Secure Multi-party Computation (MPC) in distributed settings, namely, the \textit{distributed model}, which has utility comparable to \textit{central model} while, under specific security assumptions, preventing parties from obtaining others' information. One challenge of realizing DP in \textit{distributed model} is efficiently sampling noise with MPC. Although many secure sampling methods have been proposed, they have different security assumptions and isolated theoretical analyses. There is a lack of experimental evaluations to measure and compare their performances.
We fill this gap by benchmarking existing sampling protocols in MPC and performing comprehensive measurements of their efficiency. First, we present a taxonomy of the underlying techniques of these sampling protocols. Second, we extend widely used distributed noise generation protocols to be resilient against Byzantine attackers. Third, we implement discrete sampling protocols and align their security settings for a fair comparison. We then conduct an extensive evaluation to study their efficiency and utility. Our experiments show that (1) malicious protocols based on a technique called bitwise sampling are more efficient than other methods, and using an oblivious data structure can reduce the circuit size in high-security regimes, (2) the cost of realizing malicious security is high, under the assumption of semi-honest, using a method named distributed noise generation is much more efficient, and (3) the utility loss caused by sampling noise in MPC is small, which to a certain extent eliminates utility concerns when using the DDP protocol in practice. We also open-source our code The code of our benchmark is now available at \url{https://github.com/yuchengxj/Secure-sampling-benchmark}.
\end{abstract}




\maketitle

\section{Introduction}

Differential Privacy (DP) is a strong notion of privacy-preserving algorithms~\cite{dwork2006calibrating}. DP has been widely used in many scenarios, such as the analysis of personal interest, medical analysis, and machine learning. 
The classic definition of DP assumes that a trusted central server can collect sensitive user information and then add noise to the results of specific queries, namely, the \textit{central model}. However, in situations where the server is not trustworthy, this definition raises privacy concerns. One possible solution is \textit{local model}, also known as Local DP (LDP) \cite{xiong2020comprehensive}, where each user perturbs the input locally and then sends the results to the server. 
However, the \textit{local model} usually requires significantly more samples to achieve the same utility as the \textit{central model}. 

An alternative is to take advantage of \textit{secure multi-party computation} (MPC), which enables users to jointly compute a function without revealing their inputs. By evaluating the aggregation function and then adding perturbations to the output in MPC, multiple users can obtain the final result satisfying DP without a trustworthy server.  This is also called Distributed DP (DDP). 
The main challenge to realize DDP is how to produce random noise in MPC. Recent works have proposed many \textit{sampling protocols} to efficiently sample noise from particular distributions \cite{dwork2006our, champion2019securely, goryczka2015comprehensive, bohler2021secure, anandan2015laplace,kairouz2021distributed, pentyala2022training}.

\begin{definition}[Distributed Differential Privacy] \label{TH:DDP}
A randomized protocol $\Pi_f$ implemented among 
$m$ computing parties $p = \{p_{0}, . . . , p_{m-1}\}$, satisfies Distributed Differential Privacy $w.r.t.$ a coalition $c \subset p$ of semi-honest computing parties of size $t$, if the following condition holds: for any neighbouring datasets $D$, $D'$ differing in a single entry, and any possible set $S$ of views for protocol $\Pi$,
\begin{equation}
    {\sf Pr}\left[ {\sf VIEW}^p_{\Pi}(D) \in S \right] \leq e^{\epsilon} {\sf Pr}\left[ {\sf VIEW}^p_{\Pi}(D') \in S \right] + \delta_{\kappa},
\end{equation}
where $\delta_{\kappa}$ is a negligible term associated with a security parameter $\kappa$. 
\end{definition}

\begin{figure}[t]
    \centering
    \includegraphics[width=0.4\textwidth]{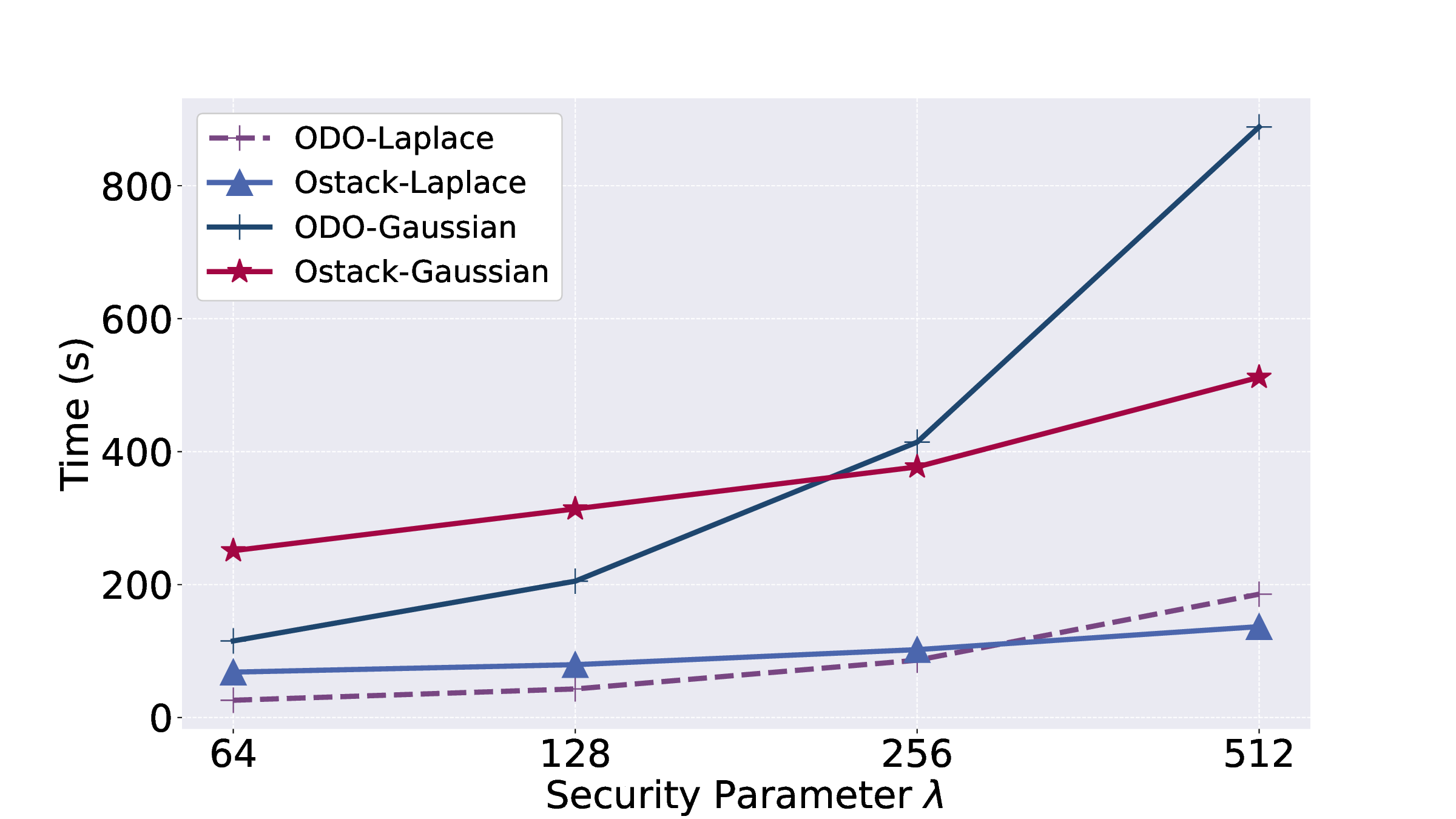}
    \caption{The running time of sampling protocols for $n=4096$ discrete Laplace (ODO-Laplace, Ostack-Laplace) and Gaussian (ODO-Gaussian, Ostack-Gaussian) samples under different security parameters $\lambda \in \{64, 128, 256, 512\}$. }
    \label{fig:intro}
\vspace{-0.5cm}
\end{figure}

The sampling protocols work as bridges between MPC and DP, significantly affecting privacy-preserving algorithms' performance in the distributed setting.  We give an example in Figure \ref{fig:intro} to show the running time of using different sampling protocols to generate $n=4096$ noise variable under security parameters $\lambda \in \{64, 128, 256, 512\}$. As we can see, the efficiency of sampling protocols varies, and an inappropriate choice of protocol can result in a long execution time in practice.
While these sampling protocols can be integrated into similar DDP pipelines, their methods vary and have different security assumptions. 
More importantly, prior evaluations of these methods were conducted in isolation, with varying settings. 
The lack of a consistent evaluation hinders researchers from understanding and selecting the most efficient sampling algorithms in diverse settings, which motivates us to benchmark different sampling protocols and conduct comprehensive evaluations. 
Our contributions are as follows.


\textbf{(1) Review and Taxonomy. }
We first review existing sampling protocols for DDP and propose a taxonomy to categorize them based on their sampling techniques into three general approaches: \textit{distributed noise generation}, \textit{uniform transformations}, and \textit{bitwise sampling}. We also analyze the security model of these protocols and identify the gap in the semi-honest \textit{distributed noise generation}.


\textbf{(2) Benchmark and Alignment. } We implement existing sampling protocols with the MP-SPDZ framework \cite{keller2020mp}. Our implementations are versatile and compatible with various underlying MPC protocols, which also work as a library, enabling practitioners to realize DDP on their aggregation results in MPC easily. 
In the benchmark, we also extend the \textit{distributed noise generation} with an additional MPC implementation of a statistical test, which checks \textit{poisoning attacks} to partial noise. 
We also create a framework that aligns the statistical distances by different sampling protocols. This guarantees protocols can achieve identical DDP protection, where we can fairly compare their efficiency.

\textbf{(3) Experimental Evaluation. }
With the benchmark, we conduct comprehensive experiments, aiming to understand the question of which is the most effective sampling protocol under different scenarios.
We vary the targeted number of samples, the security parameter, the privacy budget, and the number of computing parties to measure the efficiency of different sampling protocols. Our metric spans from empirical results like running time and communication to static indicators like the number of AND gates and random bits input from parties. We also instantiate the basic counting query on the real-world dataset to verify secure sampling protocols' utilities. 

\textbf{(4) Findings. }
Given the experimental results, we identify two interesting trends: protocols relying on bitwise sampling tend to be more efficient than alternative methods, and employing oblivious data structures can reduce the circuit size when heightened security is required. We give detailed guidelines for choosing the proper sampling protocol when considering different scenarios.
Additionally, through our empirical study, we demonstrate that noise sampled in DDP and pure CDP have almost the same utility guarantee, which to a certain extent eliminates utility concerns when using the DDP protocol in practice.

\mypara{Roadmap.}
The remainder of this paper is organized as follows. In Section \ref{sec:relate}, we introduce related works. Section \ref{sec:pre} provides the preliminary for differential privacy and secure multi-party computation. In Section \ref{sec:main}, we review existing sampling protocols. In Section \ref{sec:extend-DNG}, we analyse and extend widely used distributed noise generation protocols. In Section \ref{sec:bench}, we describe our benchmark and categorize their statistical distance with the same framework. Section \ref{sec:exp} presents evaluation results. We give some takeaways from our evaluation in Section \ref{sec:takeaway} and a conclusion in Section \ref{sec:con}.

\section{Related Work} \label{sec:relate}

\textbf{Differential Privacy. } Differential privacy (DP)~\cite{dwork2006calibrating,dwork2014algorithmic} is a strong notion preventing individuals from being inferred from the aggregated data. In the central DP (CDP) setting which assumes the data curator is trusted,  many types of data statistics methods with privacy guarantee have been proposed, including range query \cite{hay2009boosting, qardaji2013understanding}, data stream release \cite{chan2011private, wang2021continuous}, synthetic dataset generation \cite{tao2021benchmarking, zhang2021privsyn} and machine learning \cite{abadi2016deep, wei2023dpmlbench}, etc. 
Moreover, a more stringent definition named local DP (LDP)~\cite{DBLP:conf/focs/KasiviswanathanLNRS08} is widely used when the curator is untrusted. There are also many works designing LDP algorithms to protect the privacy of users against the adversarial curator \cite{zhang2018calm, ren2022ldp, cormode2018privacy, wang2017locally}. However, because LDP perturbs each user's input, the final aggregated results have poorer utility than CDP. An alternative is called the shuffle DP~\cite{cheu2019distributed, cheu2021differential}, which uses a non-colluding shuffler to permute LDP random reports from clients.  The utility in the shuffle model can be superior to the LDP setting, but it remains less optimal than CDP \cite{balcer2019separating}.

\textbf{Distributed Differential Privacy. }  
We focus on another approach that leverages secure multi-party computation (MPC).
It can achieve the same trust model as shuffle DP (introducing a non-colluding party) but can be more versatile.  It achieves better utility than shuffle DP, but with a sacrifice on computational cost.
We call this distributed DP (DDP). 
Several works have proposed practical MPC protocols to realize different types of DDP statistics, including median queries \cite{bohler2020secure, boehler2022secure}, heavy hitter \cite{bohler2021secure, boyle2019secure}, graph query \cite{roth2021mycelium}, and machine learning \cite{cheu2019distributed, pentyala2022training}. 
Sampling noise in MPC is the foundation for achieving DDP. One branch of sampling protocol is distributed noise generation \cite{anandan2015laplace, bohler2021secure, goryczka2015comprehensive, heikkila2017differentially, kairouz2021distributed}. It lets parties locally sample noise, which is then summed in MPC.  This approach can be easily and efficiently implemented.  However, it can only achieve semi-honest security, and the byzantine adversaries can use incorrect partial noise to violate differential privacy \cite{dwork2006our, keeler2023dprio, xiang2023practical}. To realize malicious security, Dwork et al. \cite{dwork2006our} propose to first securely sample biased coins and compose them into the sample from the geometric distribution. Subsequently, Champion et al. \cite{champion2019securely} construct an oblivious data structure to improve the efficiency of sampling biased coins. Wei et al.~\cite{wei2023securely} use rejection sampling to convert the discrete Laplace samples into discrete Gaussian samples, where they use the method in Dwork et al. \cite{dwork2006our} as a building block. Other transform-based protocols \cite{eigner2014differentially, pentyala2022training} require float-point non-linear arithmetic, which is significantly slower in MPC protocols. Biswas et al. \cite{biswas2018interactive} propose a verifiable mechanism sampling noise from Binomial distribution to realize DP. Due to the unbiased sampled coin, it can not directly be applied to sample noise from Laplace distribution. 

While there are different protocols using various approaches, we identify a lack of comprehensive understanding of their security guarantees and performance under different settings.  In this paper, we benchmark widely used sampling protocols to conduct evaluations.

\section{Preliminaries} \label{sec:pre}

\subsection{Differential Privacy}  \label{sec:dp}
Differential privacy (DP) is a privacy notion to protect the privacy of any specific individuals under the aggregate statistics. 

\begin{definition}[Differential Privacy \cite{dwork2006calibrating}] \label{TH:DP}
A randomized algorithm ${\rm M}: \mathbb{F} \rightarrow \mathbb{O}$ is $(\epsilon, \delta)$-differential privacy (DP) if for any pair of neighbouring datasets $D, D' \in \mathbb{F}$ that differ by a single record, and for any  possible subset $S$ of ${\rm M}$'s output,
\begin{equation}
    {\sf Pr}\left[ {\rm M}(D) \in S \right] \leq e^{\epsilon} {\sf Pr}\left[ {\rm M}(D') \in S \right] + \delta,
\end{equation}
where the parameter $\epsilon \geq 0$ denotes the privacy budget, and $\delta \geq 0$ denotes the probability that privacy is violated.
\end{definition} 

When $\delta \neq 0$, the mechanism is approximate DP ($(\epsilon, \delta)$-DP). Otherwise, we call it pure DP ($\epsilon$-DP).


\mypara{Primitives for DP.}
In order to achieve DP for numerical statistics, one can add noise to the function output. Let $\Delta$ be the \textit{sensitivity} of function $f$, which is the largest change to the function output after changing a single record. 

A common way to achieve $\epsilon$-DP is adding noise sampled from the Laplace distribution with parameter $\Delta/\epsilon$ ($\Delta$ is measured in $l_1$ distance), i.e., the Laplace mechanism \cite{dwork2014algorithmic}. 

\begin{theorem}[Laplace Mechanism \cite{dwork2014algorithmic}] \label{TH:LM}
Given a function $f$: $\mathbb{F} \rightarrow \mathbb{O}^k$, and a dataset $D \in \mathbb{F}$, the Laplace Mechanism is defined as ${\rm M}_{\mathsf{Lap}}(D, f(\cdot), \epsilon) = f(D) + (L_0, ..., L_{k-1})$, where $L_i$ are i.i.d. random variables sampled from zero-mean Laplace distribution $\Lap{\frac{\Delta}{\epsilon}}$. 
\end{theorem}

Another mechanism to achieve DP is the Gaussian Mechanism ${\rm M}_{\mathsf{N}}$, which adds noise sampled from the zero-mean Gaussian distribution $\Gau{\sigma^2}$ and the sensitivity $\Delta$ is measured by $l_2$ distance.

\begin{theorem}[Gaussian Mechanism \cite{dwork2014algorithmic}] \label{TH:GM}
Let $\epsilon \in (0,1)$. 
The Gaussian Mechanism with parameter $\sigma = \Delta\sqrt{2\ln{(1.25/\delta)}} / \epsilon$ satisfies $(\epsilon, \delta)$-DP.
\end{theorem}
There are recent improved results~\cite{balle2018improving,mironov2017renyi} about Gaussian mechanism.  In this paper, for presentation simplicity, we use the classic result.  There are also discrete versions of these mechanisms with similar privacy guarantees.

\label{sec:discrete mechanisms}

\begin{theorem}[Discrete Laplace Mechanism] \label{TH:DLM}
Given a function $f$: $\mathbb{F} \rightarrow \mathbb{O}^k$, and a dataset $D \in \mathbb{F}$, the discrete Laplace Mechanism is defined as ${\rm M}_{\mathsf{Lap}_\mathbb{Z}}(D, f(\cdot), \epsilon) = f(D) + (L_0, ..., L_{k-1})$, where $L_i$ are i.i.d. random variables sampled from zero-mean discrete Laplace distribution $\dLap{\frac{\Delta}{\epsilon}}$. And the ${\rm M}_{\mathsf{Lap}_\mathbb{Z}}$ satisfies $\epsilon$-DP.
\end{theorem}

\begin{theorem}[Discrete Gaussian Mechanism] \label{TH:DLM}
Given a function $f$: $\mathbb{F} \rightarrow \mathbb{O}^k$, and a dataset $D \in \mathbb{F}$, the discrete Gaussian Mechanism is defined as ${\rm M}_{\mathsf{N}_{\mathbb{Z}}}(D, f(\cdot), \epsilon) = f(D) + (Y_0, ..., Y_{k-1})$, where $Y_i$ are i.i.d. random variables sampled from zero-mean discrete Gaussian distribution $\dGau{\sigma^2}$. And the ${\rm M}_{\mathsf{N}_{\mathbb{Z}}}$ satisfies $(\epsilon,\delta)$-DP for
$\delta=\mathsf{Pr}_{Y\gets \dGau{\sigma^2}} \left [ Y > \frac{\epsilon \sigma^2}{\Delta}-\frac{\Delta}{2} \right ]  - e^{\epsilon} \cdot \mathsf{Pr}_{Y\gets \dGau{\sigma^2}} \left [ Y > \frac{\epsilon \sigma^2}{\Delta} + \frac{\Delta}{2} \right ]  $.
\end{theorem}

In this work, we benchmark DP protocols in MPC.  We focus on discrete protocols because those are more naturally supported in MPC, and most existing work is discrete.  That said, our library also includes continuous protocols.  
Other primitives, like Exponential Mechanism, Report Noisy Max, and Sparse Vector Technique, can be reduced to sampling Laplace or Gaussian noise.  We present their details in Appendix~\ref{sec:other dp primitive}.

\subsection{Secure Multi-party Computation}
Secure Multiparty Computation (MPC) \cite{goldreich2009foundations, yao1982protocols} allows a set of parties to jointly compute a function $y = f(D_0,..., D_{m-1})$ without revealing their inputs $D_i$ $(0\leq i \leq m-1)$. After the computation, all the parties can only know the result $y$. Currently, the two main paradigms to implement MPC are \textit{garble circuits} \cite{micali1987play, lindell2009proof} and \textit{secret sharing} \cite{bendlin2011semi, corrigan2017prio}. 
In this paper, we focus on a secret sharing scheme, which offers better scalability in the multi-party setting; our evaluation also involves garbled circuits for two-party settings.

The secret-sharing-based-MPC splits each party's input $d_i$ into $m$ pieces of shares $\langle d_i \rangle_j$ $(0 \leq j \leq m-1)$ by a \proto{Share} function. Next, each party $i$ runs the protocol with $m$ shares $\langle d_1 \rangle_i, ..., \langle d_{m-1} \rangle_i$ and gets the shares of output $\langle y \rangle_i$, which are used to reconstruct the plain text output $y$ with a \proto{Rec} function. In a $(t, m)$-secret sharing protocol, $t$ shares are sufficient to reconstruct the plain text with \proto{Rec}.  For simplicity, we use $\langle s \rangle$ to denote shares $\{\langle s \rangle_1, ..., \langle s \rangle_{m-1}\}$ of $s$ among $m$ parties. 

The adversary models of MPC protocols can be classified into the \textit{semi-honest} model and the \textit{malicious} model. The former assumes that the adversaries are curious about others' private data but still follow the protocol, even if their deviation may not be detected; the latter assumes that adversaries can deviate from the protocol arbitrarily, thus some detection/prevention scheme has to be implemented.
For more details, one can refer to~\cite{evans2018pragmatic}.

Recently, some frameworks, such as SCALE-MAMBA \cite{aly2021scale} and MP-SDPZ \cite{keller2020mp} have improved the MPC performance. In this paper, we implement existing solutions in both semi-honest and malicious settings for realizing DP in MPC using MP-SPDZ \cite{keller2020mp} framework and compare their performance. We use MP-SPDZ since it has been adopted in many MPC applications \cite{bohler2021secure, wei2023securely, humphries2022selective, dalskov2021fantastic} and supports various popular underlining protocols in both secret shares and garbled circuits.
The MPC subprotocols mentioned in this paper are listed in Appendix \ref{app:notations}. In most cases, they operate on secret shares.

\subsection{Distributed Differential Privacy} \label{sec:ddp}

The conventional definition of DP holds against computationally unbounded adversaries. In this paper, we consider distributed DP (DDP) against computationally bounded adversaries in the distributed setting \cite{eigner2014differentially}.

Specifically, in DDP, we assume there is a set of $m$ computing parties and $m'$ input parties (an input party can also serve as a computing party). Each input party $i$ gives a private input $d_i$ to the computationally bounded, untrusted and non-colluding computing parties to construct a dataset $D$. Then, computing parties jointly compute the final statistic results on $D$ by running the protocol $\Pi_f$. Let ${\sf VIEW}^p_{\Pi}(D)$ be the view of any computing party $p$ when executing the protocol $\Pi$ on dataset $D$, including all exchanged messages and internal state, and $\kappa$ be the view security parameter. The protocol $\Pi_f$ satisfies the following definition:

\begin{definition}[Distributed Differential Privacy \cite{eigner2014differentially}] \label{TH:DDP}
A randomized protocol $\Pi_f$ implemented among 
$m$ computing parties $p = \{p_{0}, . . . , p_{m-1}\}$, satisfies Distributed Differential Privacy $w.r.t.$ a coalition $c \subset p$ of semi-honest computing parties of size $t$, if the following condition holds: for any neighbouring datasets $D$, $D'$ differing in a single entry, and any possible set $S$ of views for protocol $\Pi$,
\begin{equation}
    {\sf Pr}\left[ {\sf VIEW}^p_{\Pi}(D) \in S \right] \leq e^{\epsilon} {\sf Pr}\left[ {\sf VIEW}^p_{\Pi}(D') \in S \right] + \delta_{\kappa},
\end{equation}
where $\delta_{\kappa}$ is a negligible term associated with a security parameter $\kappa$. 
\end{definition}

\mypara{General Pipeline of DDP.}  
At the high level, the end-to-end process of a DDP protocol works in four steps. 
In the process of aggregation statistic, each input party $i$ sends its $d_i$ to the untrusted but non-colluding computing parties to form a combined dataset $D = \{d_0, ..., d_{m'-1}\}$. Each $d_i$ can be a single data point (in the fully distributed setting) or a collection of samples (the input party already collects data in raw format). 
Formally, the pipeline is as follows. 

\begin{itemize}[leftmargin=*]
    \item \textit{Secret Sharing}. Each input party $i$ splits its data $d_i$ into shares $\langle d_i \rangle=\texttt{Share}(d_i)$ and sends them to $m$ computing parties. After that, the input parties can go offline.
    
    \item \textit{Statistics Aggregation}. The computing parties run secret-sharing-based protocols $\Pi_f$ of some function $f$ to get the shares of aggregation result $\langle f(D) \rangle$.
    
    \item \textit{Noise Sampling}. The computing parties run an additional secure sampling protocol $\Pi_s(x_0, ..., x_{m-1})$ (where $x_i$ is from computing party $i$) to get the shares of noises $\langle \eta \rangle$. 
    
    \item \textit{Addition and Reconstruction}. The computing parties add the noise to the aggregation result to get shares of the final output $\langle y \rangle = \langle f(D) \rangle + \langle \eta \rangle$. Then, the reconstruction function is called to reveal the differential private result $\proto{Rec}(\langle y \rangle)$. Note that $\langle f \rangle $ and $\langle \eta \rangle$ can be in vector forms.
\end{itemize}

\section{Overview of Sampling Protocols} \label{sec:main}




\begin{table*}
    \centering
    \caption{Summary of existing secure sampling methods to achieve DP with MPC. }
    \label{tab:summary}
    \begin{tabular}{c|ccc}
        \toprule
        Sampling Process & Mechanism & Adversary Model & Method \\
        \midrule
        \multirow{4}{*}{Distributed Noise Generation} & Continuous Laplace & Semi-honest & Summing Gamma Noise \cite{bohler2021secure, anandan2015laplace} \\
        & Continuous Laplace & Semi-honest & Summing Laplace Noise \cite{goryczka2015comprehensive} \\
        & Continuous Gaussian & Semi-honest & Summing Gaussian Noise \cite{heikkila2017differentially} \\
        & Discrete Gaussian & Semi-honest & Convolution \cite{kairouz2021distributed} \\
        \midrule
        \multirow{3}{*}{Uniform Transformation} & Continuous Laplace & Malicious & Inverse Transformation \cite{eigner2014differentially} \\
        & Discrete Laplace & Malicious & Inverse Transformation \cite{eigner2014differentially} \\
        & Continuous Gaussian & Malicious & Box and Muller Transformation \cite{pentyala2022training} \\
        \midrule
        \multirow{3}{*}{Bitwise Sampling} & Discrete Laplace & Malicious & ODO Sampling \cite{dwork2006our} \\
        & Discrete Laplace & Malicious & Ostack-based Sampling \cite{champion2019securely} \\
        & Discrete Gaussian & Malicious & Rejection Sampling \cite{wei2023securely} 
        \\
        \bottomrule
    \end{tabular}
\end{table*}

Now, we focus on secure sampling protocols $\Pi_s$ for the Laplace or Gaussian mechanism proposed by recent studies. 
According to their underlying sampling process, we classify these methods into three main categories: (1) \textit{distributed noise generation}, (2) \textit{uniform transformation}, and (3) \textit{bitwise sampling}.
Table \ref{tab:summary} shows a summary of methods based on this categorization (along with two more dimensions of what DP mechanism they support and the adversary model in MPC).  
In what follows, we review these approaches in more detail and discuss the limitations of semi-honest protocols.

\subsection{Semi-honest Sampling Protocols}\label{sec:semi-dng}
Distributed noise generation (DNG) is the simplest way to add DP noise in MPC.  In this approach, each party locally samples partial noise and sends it to the MPC protocol $\Pi_s$. 
These partial noises can be combined to obtain the required distribution. The combining process usually requires only ADD operations in MPC, which is highly efficient. 
Below, we briefly review the existing works using DNG protocols for Laplace and Gaussian noise.

\mypara{DNG for Gaussian Noise \cite{heikkila2017differentially, kairouz2021distributed, dwork2006our}. }The distributed sampling for Gaussian noise is direct since the sum of Gaussian variables is also a Gaussian variable. Thus the generation of $Y \sim \Gau{\sigma^2}$ can be achieved by $Y = \sum_{i=0}^{m-1} Y_i$, where $Y_i$ is drawn by party $i$ with $Y_i \sim \Gau{\frac{\sigma ^ 2}{m}}$ \cite{heikkila2017differentially, kairouz2021distributed}. 
In our evaluation, we assume that the summation of discrete Gaussian samples is a Gaussian sample since they have been proven extremely close \cite{kairouz2021distributed}.  

\mypara{DNG for Laplace Noise \cite{goryczka2015comprehensive, bohler2021secure}. }
Different from Gaussians, Laplacians do not add up to Laplacian. 
However, distributed parties can generate other kinds of noise that add up to a Laplace noise.  In particular, partial noises from the Negative Binomial distribution $Y_{i} \sim  \mathsf{NB} (1/m, e^{-\frac{\Delta}{\epsilon}})$ can add up to a geometric noise.  Then, an additional unbiased bit is used to convert them to discrete Laplace samples.
Goryczka et al. \cite{goryczka2015comprehensive} summarized the partial noise to generate continuous Laplace and geometric samples. Böhler et al.\cite{bohler2021secure} also apply distributed noise generation to compute heavy hitters.

We summarize the partial noise generated by computing parties locally and arithmetics/operations to obtain target noise in MPC in Table~\ref{tab:summary-dng} in Appendix \ref{app:dng}. 

{
There is a privacy risk in DNG protocols, namely, the colluded attack. The corrupt parties can collude with each other to "subtract away" their noise samples from the final result and reduce the noise to below the amount needed for the desired differential privacy guarantee. Therefore, for example, if $50\%$ parties collude, each party needs to add additional (twice as much) noise as in the central model because up to half of the parties may be corrupt and subtract their portion of the noise. For the original partial Gaussian noise with variance $\sigma^2$, when there are $\alpha$ proportion of colluded parties, each party should provide Gaussian noise with variance $\hat{\sigma} ^ 2$,
\begin{equation} \label{eq:new_variance}
    \hat{\sigma} ^ 2 = \sigma ^ 2 / (1-\alpha)
\end{equation}

Such additional noise will cause a reduction of utility for the final results. We measure this impact in Section~\ref{exp:collude}.
}

\subsection{Poisoning Attacks to Semi-honest Protocols} \label{sec:semi-malicious}

The most significant limitation of semi-honest protocols is that they initially provide no guarantee when parties change their input arbitrarily, i.e., poisoning attack. It can result in incorrect aggregated noise to destroy privacy or utility.
As for the computing parties cooperatively sampling noise, there are two types of attacks in the presence of malicious computing parties. Also, there is a possible attack named data poisoning attacks, which is caused by the malicious input parties.


\mypara{Zero Noise.} It is possible that an adversarial server keeps providing $0$ as the partial noise, which causes the violation of the desired guarantee for privacy since the protocol eventually adds noises with smaller variance to the outputs.
    One possible solution is that each computing party provide a larger magnitude of noise.
    Taking the DNG protocol for the Gaussian mechanism proposed by Dwork as an example \cite{dwork2006our}, $m$ computing parties can sample partial Gaussian noise with variance $\frac{3\sigma^2}{2 m}$, which can still realize required DP in the setting of at least $\frac{2}{3}$ honest servers. The required number of honest computing parties can be reduced to one if each party provides a noise with variance $\sigma^2$. Although this method provides a privacy guarantee while resulting in an expected error of $m\sigma^2$ \cite{keeler2023dprio}.

\mypara{Sampling Larger Amount of Noise.} Another poisoning attack is that the malicious computing party generates a significant amount of noise as the partial noise input to the MPC protocol. Due to the property of MPC, this illegal input can not be viewed by other participants. Moreover, the numerical noise from parties is unbounded and directly added to the numerical answer, significantly decreasing the protocol's utility. Although \cite{xiang2023practical} uses non-MPC methods, applying statistical tests can check whether inputs from computing parties fall in some interval and whether they can pass the null hypothesis that they are sampled from the specific distribution, these methods can not be directly applied to the MPC setting since the noise are no longer accessible to be tested. Otherwise, the statistical test in MPC needs to be designed.

{

\mypara{Data Poisoning Attack.}
The data poisoning attack also exists in the \textit{Statistic Aggregation} phase of the  DDP pipeline. The input parties can manipulate the inputs to skew the query results before the \textit{Noise Sampling} phase~\cite{biswas2018interactive}, which is out of the scope of secure sampling protocols. Next, we discuss such malicious settings and corresponding defense approaches other than secure sampling.

The secure aggregation in federated learning averages gradient across multiple local models, where the DP sampling can be further used to provide output protection~\cite{xiang2023practical}. However, the malicious client can input large gradients to skew the aggregated gradients. To this end, the zero-knowledge proof (ZKP) is often used by the computing parties, which checks whether the inputs satisfy $L_1, L_2$ and $L_{\infty}$ bounds. One direct solution is the Bulletproof protocol for bound checking~\cite{bunz2018bulletproofs}. There is a sequence of studies on checking gradient bound in federated learning~\cite{bell2023acorn, bell2020secure, roy2022eiffel}. Such strategies can also be applied in other scenarios requiring mean statistics.

The adversary in DDP statistics queries is similar to those in LDP settings, where each input party provides invalid inputs or dishonest responses. For example, the DDP heavy hitter collects each user's binary response~\cite{bohler2021secure}. With ZKP, each input is proved to be in $\{0,1\}$. Thus, a coalition of $t$ malicious parties only changes each aggregated count of values at most $t$, which can be further mitigated by the idea of normalization by observing consistency~\cite{wang2019locally}. Also, in the DDP median, the malicious inconsistency of inputted value-rank pairs should be checked~\cite{bohler2020secure}. Note that the defense strategies for the data poisoning attack can only mitigate the effect of poisoning attacks instead of eliminating them since the inputs are determined by clients.

}

There are standard ways to transform semi-honest to malicious security~\cite{evans2018pragmatic}, but they focus on the process of computation rather than inputs, assuming data is deterministic and cannot test random variables.
We propose a method based on the Kolmogorov–Smirnov (KS) test later in Section \ref{sec:extend-DNG}.  The high level of our KS test is to reject the hypothesis that the generated samples come from a specific distribution when the results are poisoned.
A common way to realize malicious for the randomness is using XOR. The secure sampling protocols for DP with XOR  originated from the protocol proposed by Dwork et al.~\cite{dwork2006our} (we call it the ODO protocol in the following). 
The idea is simple: XORing the input bits from each party can obtain randomness as long as at least one party is inputting a truly random bit.  Compared to previous methods, this is extremely simple but much more efficient. Still, the drawback is this only gives a random bitstring, and we need to design sampling protocols to transform randomness into specific distributions.


\subsection{Malicious Sampling Protocols}

\mypara{Uniform transformation. }
Given the random bitstring, each from $\{0, 1\}$, constructed from XOR described above. 
Next, the bitstring can be viewed as a uniformly random number $u \sim \mathsf{U}(0, 1)$. Next, 
the most straightforward idea is to compare it with a `distribution table' from the target distributions.  That is, one can pre-compute the cumulative density function (CDF) $F(x)$ of the target distribution and then compare it with a uniformly generated random number $u\in \mathsf{U}(0,1)$ converted from the XORed random bitstring. If the $u$ lies between $F(x_i)$ and $F(x_{i+1})$, the sample corresponds to the value $x_{i+1}$.  However, in MPC, we have to traverse $F$ and perform comparison to prevent sensitive data leakage, which can be slow, especially when $F$ is a very detailed CDF.  Note that one can also compute $F^{-1}(u)$ with a logarithm arithmetics to get a Laplace sample \cite{eigner2014differentially}, where $F^{-1}$ is the inverted CDF of the exponential distribution.



\mypara{Bitwise sampling. } 
Dwork et al. \cite{dwork2006our} proposed a protocol (we call it \textit{ODO-sampling}) based on their observation that each bit in the geometric sample can be sampled independently. 
Based on ODO, Champion et al.~\cite{champion2019securely} proposed to use an oblivious stack data structure to obliviously pop and push bits (we call it \textit{Ostack-sampling}), which avoids complete iteration over the binary expansion of bias to hide the access pattern thus improve efficiency.
The procedure of ODO-sampling and Ostack-sampling is shown in Appendix \ref{app:sampling}.
With many biased coins from $\mathcal{B}(p)$ from different biases $p$, we can get the geometric distribution by concatenating them.
Then, the two-sided discrete Laplace sample can be obtained by transforming the one-side geometric sample.  Compared to the uniform transformation mentioned above, this approach is typically more efficient but relies on key observations (bits can be sampled independently).
Finally, we can get discrete Gaussian samples by applying rejection sampling to Laplace samples \cite{wei2023securely}.


\subsection{Statistical Distances Caused by MPC}

Our benchmark focuses on implementing discrete sampling protocols and aligning their security demand. Sampling from the continuous distribution relies on the fixed point or float point representation in MPC, and the statistical distance between the actual continuous distribution and that with a finite number representation is hard to measure.  In fact, existing works~\cite{bohler2021secure} using continuous noise in MPC did not formally quantify additional statistical distance. In this paper, we focus on discrete noise generation in MPC.

The components of statistical distance in discrete sampling protocol also vary, and incorrectly setting protocol parameters will result in a loss of security. 
Thus, we summarize the allocation of security parameter $\delta_{\kappa}$. The sources of extra statistical distance in sampling discrete noise are $\delta_t$ caused by truncating ideal distribution from $\mathbb{O} \in \mathbb{Z}$ to $\mathbb{O} \in (-N, N) \cap \mathbb{Z}$, where $N \in \mathbb{N}$ control the truncating range since MPC operates in fixed-length integers, $\delta_b$ caused by representing the bias $p$ in a finite number when sampling biased coins, $\delta_r$ from the potential failure of rejection sampling, and $\delta_p$ from the potential failure of filling the Ostack.
Note that existing works \cite{champion2019securely, wei2023securely} have analysed this for ODO-Gaussian and Ostack-Laplace, while there is no such theoretical analysis for others. We expand this landscape for all the protocols in our benchmark. More details about how to set different security parameters are given in Section~\ref{sec:align}.

\section{Verifying DNG protocol}
\label{sec:extend-DNG}


\begin{algorithm}[t]
\caption{\textbf{Check}-$\mathsf{Lap}_{Z}$ } 
\label{alg:test}
\begin{algorithmic}[1]
\INPUT Length of aggregated sample $n$, shares of aggregated noise $\langle y_1 \rangle, ..., \langle y_{n} \rangle$.
\OUTPUT The shares of Boolean variable specifying whether to reject the null hypothesis $\langle b \rangle$.
\PRECOMPUTE CDF table $\langle F[j] \rangle = \langle \mathcal{F}_{\mathsf{Lap}}(j-N) \cdot n \rangle$ for $j=1,...,2N-1$, KS distance $\langle p'_a \rangle = \langle c(a) \cdot n \cdot  \sqrt{\frac{n+2N+1}{n\cdot (2N+1)}} \rangle$ of  significance value $\alpha$.
\STATE Initialize array $\langle obs \rangle$ of size $n$.
\STATE Initialize $\langle d \rangle \gets \langle 0 \rangle$

\FOR{$i\gets 1$ \textbf{to} $n$}
    \FOR{$j \gets 1$ \textbf{to} $2N-1$} \label{algl:begintable}
        \STATE $\langle e  \rangle \gets \textbf{EQ}(\langle y_i \rangle, \langle j-N \rangle)$
        \STATE $\langle obs[i] \rangle \gets \textbf{MUX}(\langle e \rangle , \langle F[j] \rangle, \langle obs[i] \rangle )$
    \ENDFOR \label{algl:endtable}
\ENDFOR

\STATE $\textbf{SORT}( \langle obs \rangle)$

\FOR{$i \gets 1$ \textbf{to} $n$}
\STATE $\langle t \rangle \gets \textbf{ABS}(\textbf{SUB}(\langle i \rangle, \langle obs[i]\rangle))$
\STATE $\langle d \rangle \gets \textbf{MUX}(\textbf{LE}(\langle d \rangle, \langle t \rangle), \langle d \rangle, \langle t \rangle )$
\ENDFOR
\STATE $\langle b \rangle \gets \textbf{LE}(\langle p' \rangle, \langle d \rangle)$
\RETURN $\langle b \rangle$
\end{algorithmic}
\end{algorithm}

The poisoning attacks mentioned in Section~\ref{sec:semi-malicious} deviate the generated noise from its original distribution. One way to constrain the probability distribution deviation is using the Kolmogorov–Smirnov test (KS test) \cite{an1933sulla} to check the partial noise from the computing parties or the aggregated noise. The definition of the two-sample KS test is as follows.

\begin{definition} [Two-sample KS test \cite{berger2014kolmogorov}]\label{TH:ks-test}

Given empirical distribution functions $F_a$ and $F_b$ of samples $a$ and $b$ with length $n_a$ and $n_b$, Kolmogorov–Smirnov test is defined by
\begin{equation*}
    D_{a,b} = \sup_x |F_a (x) - F_b (x)|. 
\end{equation*}
The null hypothesis that $a$ and $b$ are sampled from the same distribution is rejected at a significance level $\alpha$ if 
\begin{equation*}
    D_{a,b} > c(\alpha) \sqrt{\frac{n_a+n_b}{n_a \cdot n_b}},
\end{equation*}
where $c(\alpha) = \sqrt{-\ln(\frac{\alpha}{2}\cdot \frac{1}{2})}$, and the significance level $\alpha$ is the probability of rejecting the null hypothesis when the null hypothesis is true.

\end{definition} 

Applying the KS test framework in our setting (assuming discrete Laplace; discrete Gaussian naturally follows), we hold the null hypothesis that the aggregated noises $y=y_0, ..., y_{n-1}$ are sampled from the discrete Laplace distribution $\Lap{\frac{\Delta}{\epsilon}}$. The KS test computes its empirical Cumulative Distribution Function (eCDF) by $F_{obs}(y')=\frac{1}{n} \sum_{i=0}^{n-1}\textbf{1}_{y_i<y'}$, which is the number of samples smaller than $y'$ in samples $y$. 
Next, the KS statistics $D_{obs, \mathsf{Lap}}={\rm sup}_{y'} |F_{obs}(y') - F_{\mathsf{Lap}}(y')|$ is computed, where the second sample $F_{\mathsf{Lap}}$ is the true CDF of $\Lap{\frac{\Delta}{\epsilon}}$. Assuming that the discrete Laplace samples are truncated into $(-N, N) \cap \mathbb{Z}$, where $N$ denote the range of finite representation of integers in DNG protocol, we can precompute a $F_{\mathsf{Lap}}$ of length $2N-1$. 
Then, fixing the significance level $\alpha$, if $D_{KS}>c(\alpha) \sqrt{\frac{n+2N-1}{n\cdot (2N-1)}}$, we can reject the null hypothesis that the aggregated samples are not from $\Lap{\frac{\Delta}{\epsilon}}$ at the significance level of $\alpha$. In this paper, we set $\alpha=0.05$, a widely adopted significance level. 
Note that a similar method is proposed in \cite{du2023dp}. Their check is applied to the sum of noise and gradient sent by participants, aiming to find whether they are from the same distribution, whereas we directly check the noises in MPC and find whether they are from the specific distribution.

We implement the procedure of the one-sample KS test in MPC protocol \texttt{Check}, as shown in Algorithm~\ref{alg:test}. 
We assume all the partial noises and generated noise are truncated into $(-N, N) \cap {\mathbb{Z}}$.
After the \texttt{Secure Aggregation}, an additional step \texttt{Check} is performed on the aggregated noise samples. 
A table $F$ of rank based on the CDF of each element $j \in [2N-1]$, i.e. $F_{\mathsf{Lap}}(j-N) \cdot n$ is precomputed. The $p'_{\alpha} = c(\alpha) \sqrt{\frac{n+2N+1}{n\cdot (2N+1)}} \cdot n$ to be used for comparison is also precomputed.
Here we use $F_{\mathsf{Lap}}(j-N) \cdot n$ instead of $F_{\mathsf{Lap}}(j-N)$ because the prior can be directly compared to the rank $i$ of the element, avoiding additional divination in MPC to compute $i/n$ and compare it with $F(x_i)$. 
For each aggregated sample $y_i$. we iterate over $F$ to find its rank in a cumulative probability $\langle F[y_i] \rangle$ and store it in location $i$ of list $\langle obs[i] \rangle$ (Lines \ref{algl:begintable}-\ref{algl:endtable}). 
Subsequently, we sort $\langle obs[i] \rangle$ and compute the $\langle D \rangle = {\rm sup}_i |\langle i \rangle - \langle obs[i] \rangle|$. 
Finally, we check the validity of $\langle D \rangle$. If $\langle D \rangle < \langle p'_{\alpha} \rangle$, we reject the null hypothesis that the aggregated samples are not from $\Lap{\frac{\Delta}{\epsilon}}$.

We use the two sample KS test to constraint the KS distance between the aggregated noise and targeted noise distribution $D_{obs,\mathsf{Lap}}={\rm sup}_{y'} |F_{obs}(y') - F_{\mathsf{Lap}}(y')|$. We can bound $D_{obs,\mathsf{Lap}}$ with probability $\alpha$, the set significance level. We implement the protocol \texttt{Check} in our benchmark to mitigate the poisoning attacks on DNG-based protocols. We compare DNG with \texttt{Check} and other malicious sampling protocols in Section~\ref{exp:compare}.
Note that the KS-test is generic and can check any distribution. Therefore, it can also be applied to check the validity of each partial noise and trace back the malicious participants. Although this is slower than only checking the aggregated noise, it is currently not supported by other malicious sampling protocols. Since \texttt{Check} relies on traversing ordered arrays, its efficiency can also be improved by secure merging protocols~\cite{falk2023linear}, i.e., the equivalent values in ordered arrays $\langle y_1 \rangle, ... \langle y_n \rangle$ and $1, ..., 2N-1$ can be grouped together to derive $\langle obs\rangle$, thus reduce the complexity from $O(nN)$ to $O(n+N$).







\section{Benchmark and Parameter Alignment} \label{sec:bench}

In this section, we first introduce our benchmark to evaluate the performance of secure sampling protocols. This benchmark includes eight MPC sampling protocols for generating samples from \textit{discrete Laplace} and \textit{discrete Gaussian} distributions. We focus on the discrete sampling protocols since their samples can be compared with the `perfect' discrete samples by the statistical distance \cite{champion2019securely, wei2023securely}. Therefore, we can fix the $\delta_{\lambda}$ and compare their efficiency. Then, we give details of the relationship of security parameter $\lambda$ other parameters determining the efficiency of the protocol.

\subsection{Benchmark Protocols}

We implemented the following eight sampling protocols: the first six are the bitwise sampling protocols. DNG-Laplace and DNG-Gaussian are DNG-based methods with \texttt{Check}. The transformation-Laplace is a uniform-transformation-based protocol.
\begin{itemize}[leftmargin=*]
\item
{ODO-Laplace \cite{dwork2006our}}. It uses the direct ODO-Sampling protocol (see Appendix \ref{app:sampling}, the protocol with the smallest number of AND gates in \cite{dwork2006our}) to produce Bernoulli samples and compose them into discrete Laplace samples. Note that generating biased bits is the basic block for bitwise sampling.
\item
{Ostack-Laplace \cite{champion2019securely}}. It generates Bernoulli samples using Ostack-sampling in Appendix~\ref{app:sampling}, which includes two types of operation on an oblivious data structure as stacks. It supports \proto{RPOP} operation to obliviously produce bits in the binary expansion of the bias $p$, and \proto{CPUSH} to obliviously save the accepted biased coin.

\item{Ostack-Laplace* \cite{champion2019securely}}. An improved version of Ostack-Laplace saving the number of AND gates in \texttt{RPOP} \cite{champion2019securely}. That is, when $\epsilon$ is in the form of $2^{-i}\ln{2}, i=0,1,2,...$, one can predefine a periodic binary sequence to give the bits in $p$ in Ostack-sampling. We implement this version and always use the largest $2^{-i}\ln{2}$ that is smaller than the required $\epsilon$ as an approximation to give a slightly tighter guarantee than $\epsilon$-DP. 

\item{ODO-Gaussian \cite{dwork2006our, wei2023securely}}. This is the implementation of \cite{wei2023securely}, which uses rejection sampling to discard a portion of the samples drawn from the discrete Laplace distribution. The ODO sampling produces all the biased bits in this protocol.

\item{Ostack-Gaussian \cite{champion2019securely, wei2023securely}}. We fill the gap between the Ostack-sampling and sampling protocol for Gaussian \cite{wei2023securely} by integrating Ostack-sampling into all the procedures that generate biased coins. We also adjust the parameters (see Table \ref{table:security-param}). We allocate a fraction of security parameter $\lambda$ to compute the number of operations in Ostack-sampling, which bounds the probability of failing to produce the required number of coins.

\item{DNG-Laplace \cite{goryczka2015comprehensive}}. We implement the DNG in \cite{goryczka2015comprehensive}, which first generates partial noise from negative binomial distribution locally in input parties to form discrete Laplace noise. Since the above bitwise sampling method can achieve malicious security, We perform additional \texttt{Check} with KS tests in MPC to validate whether the aggregated noise is from the Laplace distribution, which limits the Byzantine attacks from active adversaries. 

\item{DNG-Gaussian \cite{kairouz2021distributed}}. Similarly,  we implement the DNG in \cite{kairouz2021distributed}, in which parties generate partial noises from Gaussian distribution locally. The noises are then simply summed in the data aggregation phase. We also perform the KS test in MPC to check whether the generated noise is from the Gaussian distribution. 

\item{Transformation-Laplace  \cite{eigner2014differentially}}. This protocol is the sampling protocol for the Geometric distribution in \cite{eigner2014differentially}, which conducts uniform sampling with the XORing technique and then transforms the uniform variable $u \in (0,1]$ into the inverted CDF $F^{-1}(u)$ of Exponential distribution. This protocol is clearly suboptimal since it requires expansive logarithm computation in MPC.

\end{itemize}

Our benchmark\footnote{\href{https://github.com/yuchengxj/Secure-sampling-benchmark}{https://github.com/yuchengxj/Secure-sampling-benchmark}} implemented by MP-SPDZ can also act as a library for the designers of MPC protocols. With the aggregated statistics obtained in MPC, say in secret shares, one can call our library with only one line of code to generate Laplace and Gaussian noise in MPC and add to the results, which follows the pipeline mentioned in Section \ref{sec:ddp} to achieve distributed differential privacy. 
{
Our benchmark is compatible with all the underlying target use case protocols in MP-SPDZ (including binary circuits like BMR, garbled circuits, and arithmetic circuits SPDZ and Shamir) as long as they support basic secure bit operations like AND and XOR. 
}

\subsection{Parameter Alignment}
\label{sec:align}
We give details about the sources of statistical distance in all the protocols into four parts, summarized in Table \ref{table:security-param}. The Statistical distance is due to imperfect sampling of MPC and must be quantified (resulting in increased $\delta$ in DP, as shown in Theorem ~\ref{TH:ODOLap}. The statistical distance is defined as follows, which is also called total variation distance.

\begin{definition}[Statistical Distance]
Let $\mathcal{V}$ and $\mathcal{W}$ be the probability distributions over $\mathbb{F}$. The statistical distance between $\mathcal{V}$ and $\mathcal{W}$ is defined by
\begin{align*}
    {\rm{SD}}(\mathcal{V},\mathcal{W})\triangleq\frac12 \sum_{x\in\mathbb{F}} |f_\mathcal{V} (x)- f_\mathcal{W}(x)|
\end{align*}
\end{definition}

\label{sec:stat-dist}
\begin{table*}[t]
\centering
\caption{The relationship of statistical distance $\delta_{\lambda}$ and the parameters of sampling protocols, with $\delta_{\lambda}=2(e^{\epsilon}+1)(\delta_t + \delta_b +  \delta_r + \delta_p)$ for all the protocols, where $\delta_{t}$ is caused by truncation, $\delta_b$ is caused by sampling biased coins, and the potential fail of rejection sampling and \proto{CPUSH} cause $\delta_r$ and $\delta_p$. We assign $\delta_{\lambda}$ equally to the non-zero terms for each protocol in our benchmark. }
\label{table:security-param}
\resizebox{\textwidth}{!}{
\begin{tabular}{c |  c c c c c |  c c c }
\toprule
\diagbox{$\delta_{\lambda}$}{$\Pi{s}$} & ODO-Laplace & Ostack-Laplace & Ostack-Laplace*  & DNG-Laplace & Trans-Laplace & ODO-Gaussian & Ostack-Gaussian & DNG-Gaussian \\
\midrule
$\delta_t$ & $\frac{2n\cdot e^{-\epsilon (N-1)/\Delta}}{e^{\epsilon/\Delta}+1}$ & $\frac{2n\cdot e^{-\epsilon (N-1)/\Delta}}{e^{\epsilon/\Delta}+1}$ & $\frac{2n\cdot e^{-\epsilon (N-1)/\Delta}}{e^{\epsilon/\Delta}+1}$ & $\frac{2n\cdot e^{-\epsilon (N-1)/\Delta}}{e^{\epsilon/\Delta}+1}$ & $\frac{2n\cdot e^{-\epsilon (N-1)/\Delta}}{e^{\epsilon/\Delta}+1}$ & $2n e^{-\frac{N^2 \epsilon^2 }{4\sigma ^ 2 \ln{(1.25/\delta)}\Delta^2}}$ & $2n e^{-\frac{N^2 \epsilon^2 }{4\sigma ^ 2 \ln{(1.25/\delta)}\Delta^2}}$ & $2n e^{-\frac{N^2 \epsilon^2 }{4\sigma ^ 2 \ln{(1.25/\delta)}\Delta^2}}$\\ 
\midrule
$\delta_b$ & $n(\kappa+1) 2^{-l}$ &  $n(\kappa+1) 2^{-l}$ &  $n(\kappa+1) 2^{-l}$ & 0 & $n 2^{-l}$ &  $\frac{n}{p^{*}}(2\kappa+m+2) 2^{-l}$ & $\frac{n}{p^{*}}(2\kappa+m+2) 2^{-l}$ & 0  \\ 
\midrule
$\delta_r$ & 0  & $0$ & $0$ & $0$ &  $0$ & $e^{\frac{-2(n'p^*-n)^2}{n'}}$ & $e^{\frac{-2(n'p^*-n)^2}{n'}}$ & 0\\ 
\midrule
$\delta_p$ & 0 & $\kappa \lceil \frac{n}{g} \rceil e^{\frac{-2(\frac{u}{2}-(g-1))^2}{u}}$ & $\kappa \lceil \frac{n}{g} \rceil e^{\frac{-2(\frac{u}{2}-(g-1))^2}{u}}$ & 0 & 0 & 0 & $(\kappa+m)\lceil \frac{n}{g} \rceil e^{\frac{-2(\frac{u}{2}-(g-1))^2}{u}}$ & 0 \\

\bottomrule
\end{tabular}
}
\end{table*}

\textbf{Truncation} ($\delta_t$). The statistical distance is caused by truncating the targeted discrete sample into $(-N, N) \cap \mathbb{Z}$. Since we use a finite number of bits to represent the generated noise sample, all eight protocols have $\delta_t$ in their statistical distance. The statistical distance between the truncated discrete Gaussian $\tdGau{\frac{2\sigma ^ 2 \ln{(1.25/\delta)}}{\epsilon^2 }}$ and discrete Gaussian $\dGau{\frac{2\sigma ^ 2 \ln{(1.25/\delta)}}{\epsilon^2 }}$ is $2n e^{-\frac{N^2 \epsilon^2 }{4\sigma ^ 2 \ln{(1.25/\delta)}\Delta^2}}$ \cite{wei2023securely}. The statistical distance between truncated discrete Laplace $\tdLap{\frac{\Delta}{\epsilon}}$ and discrete Laplace $\dLap{\frac{\Delta}{\epsilon}}$ is $\frac{2n\cdot e^{-\epsilon (N-1)/\Delta}}{e^{\epsilon/\Delta}+1}$ \cite{canonne2020discrete}. Note that we generate samples of length $\kappa$, thus, we have $N=2^{\kappa}+1$.

\textbf{Representing Bias} ($\delta_b$). There is a need to represent the bias $p$ with a binary expansion of length $l$  when sampling biased coins in ODO-Sampling, Ostack-Sampling and Transformation-Laplace. ODO-Laplace, Ostack-Laplace and Ostack-Laplace* all generate samples from the geometric distribution of length $\kappa$ to approximate discrete Laplace, and the total number of samples is $n$, thus they all have $\delta_b=n(\kappa+1)2^{-l}$, which have been proven in \cite{wei2023securely}. Moreover, the number of additional biased coins for rejection sampling is $m$, resulting in $\delta_b=\frac{n}{p^{*}}(2\kappa+m+2) 2^{-l}$, where $p^*$ is the acceptance rate in rejection sampling \cite{wei2023securely}. As for Trans-Laplace transforming the uniform variable $u \in (0,1]$ into the inverted CDF $F^{-1}(u)$, the statistical distance between a fix-point number $u$ in MPC protocol and $u \in \mathbb{R}$ is $2^{-l}$. Therefore, sampling $n$ fix-point number $u$  has a total statistical distance of $n2^{-l}$.

\textbf{Rejection Sampling} ($\delta_r$). 
\cite{wei2023securely} proposes to use rejection sampling to convert discrete Laplace samples to discrete Gaussian samples. Given the targeted number of discrete Gaussian samples $n$ and actual accept rate $p'_*$, the number of required Laplace samples $n'$ should be set larger than $n$ to constrain the probability of failing to generate enough Gaussian samples after rejection sampling. The relationship between $\delta_r$, $p'_*$, $n$ and $n'$ should be $\delta_r = e^{\frac{-2(n'p^*-n)^2}{n'}}$.

\textbf{Filling Ostack} ($\delta_p$). For the protocols using Ostack to sample biased coins, there is a probability that $u$ times of \textbf{CPUSH} operation does not fill the Ostack of size $g$. Formally, this probability is $\delta_p = M \cdot e^{\frac{-2(\frac{u}{2}-(g-1))^2}{u}}$, where $M$ is the number of calls to Ostack \cite{champion2019securely}. For Ostack-Laplace and Ostack-Laplace*, using Ostack of size $g$ to sample biased coin have $M=\kappa \lceil \frac{n}{g} \rceil$, since we need to generate $n$ geometric samples, and each sample has a binary expansion of length $\kappa$ \cite{champion2019securely}. As for Ostack-Gaussian, we have $M=(\kappa + m) \lceil \frac{n'}{g} \rceil$ because for each Laplace sample, we need to generate $\kappa$ biased coins to represent the Geometric sample and $m$ biased coins to perform rejection sampling.

Our benchmark fixes $\delta_t + \delta_b + \delta_r + \delta_p=2^{-\lambda}$ and equally assigns $2^{-\lambda}$ to all the non-zero terms of sampling the protocol (e.g $\delta_t + \delta_b + \delta_r=\frac{2^{-\lambda}}{3}$ for ODO-Gaussian). Then, we derive the parameters in protocols. As a result, their security guarantees are aligned, and we can compare their efficiency by running time, communication, and number of AND gates. 
Here, we describe the Distributed Differential Privacy achieved by the benchmarked discrete sampling protocols,  following the definition of DDP (Definition~\ref{TH:DDP}). The proof can be found in~\cite{wei2023securely}.

\begin{theorem} \label{TH:ODOLap}
If the targeted discrete mechanism $M$ is $(\epsilon, \delta)$-DP, then this mechanism realized by sampling protocol $\Pi_s$ is $(\epsilon, \delta + \delta_{\lambda})$-DP
,where $\delta_{\lambda}=2(e^{\epsilon}+1)(\delta_{t}+\delta_{b} + \delta_r + \delta_p)$.


\end{theorem}

\section{Experimental Evaluation} \label{sec:exp}


\begin{figure*}[!t]
    \centering

    \begin{subfigure}[b]{0.99\textwidth} 
    \includegraphics[width=\textwidth]{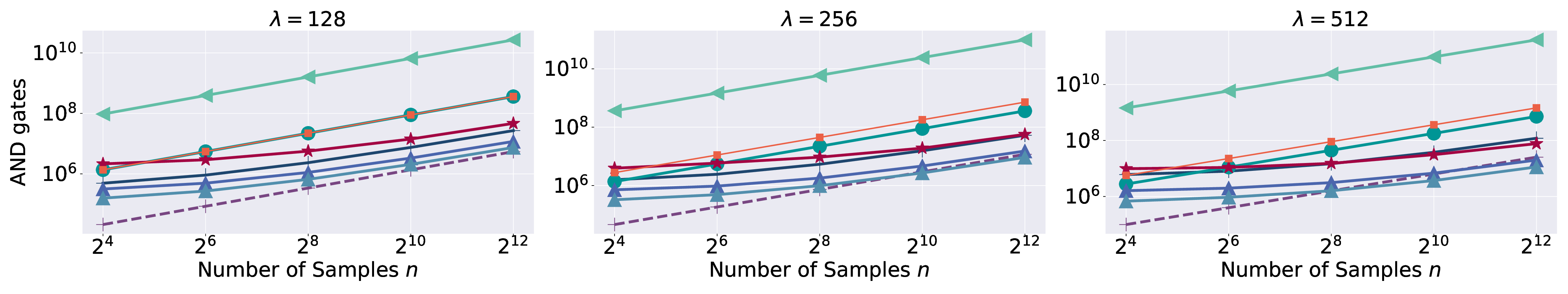}
        \caption{Number of AND gates \label{exp:and}}
    \end{subfigure}
    \begin{subfigure}[b]{0.99\textwidth} 
        \includegraphics[width=\textwidth]{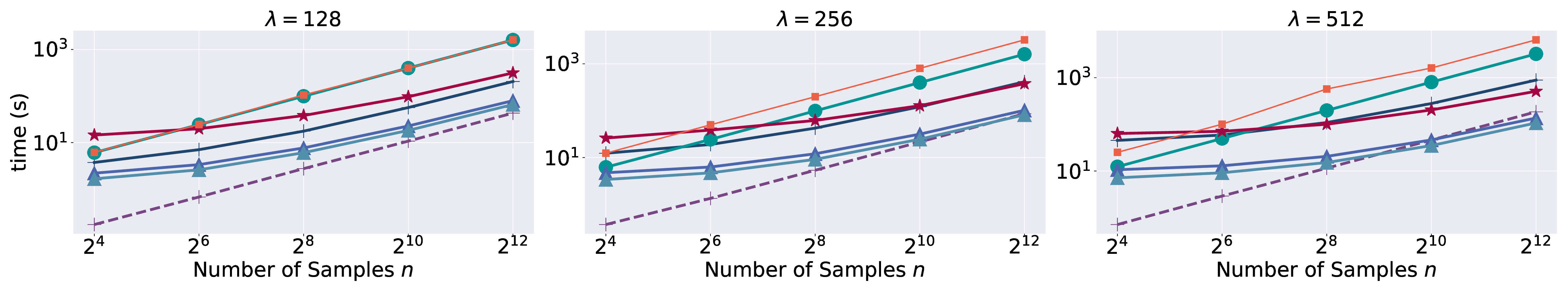}
        \caption{Running time \label{exp:time}}
    \end{subfigure}
    
    \begin{subfigure}[b]{0.99\textwidth} 
        \includegraphics[width=\textwidth]{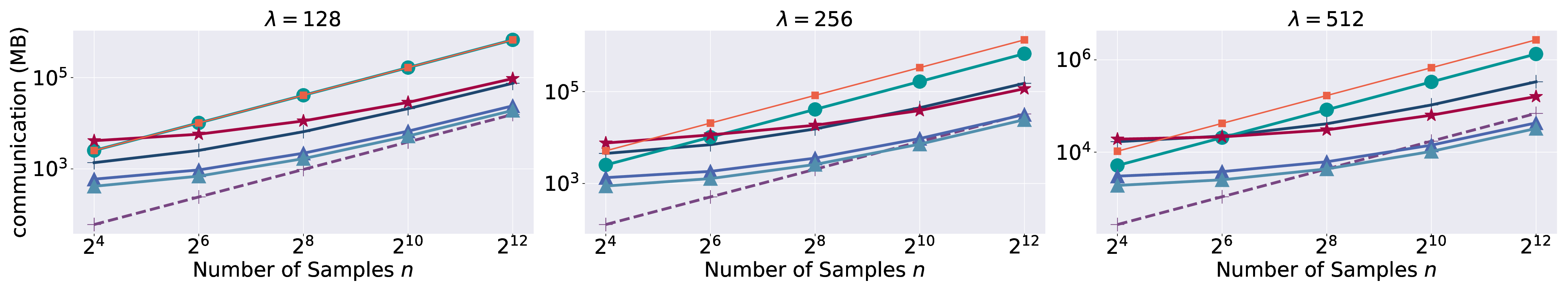}
        \caption{Communication \label{exp:com}}
    \end{subfigure}

    \begin{subfigure}[b]{0.99\textwidth} 
        \includegraphics[width=\textwidth]{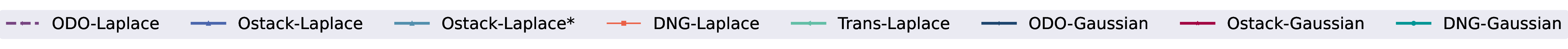}
    \end{subfigure}

    \caption{Overview of sampling protocols’ number of AND gates, running time, communication in Shamir-BMR under different $\lambda \in \{128, 256, 512\} $ and $n \in \{2^4, 2^{6}, 2^8, 2^{10}, 2^{12} \}$.}
    \label{fig:exp1_bit_and}
\end{figure*}

\begin{table*}[t]
    \centering
    \caption{Sampling protocols’ number of random bits under different $n$ and $\lambda$}
    \resizebox{0.82\textwidth}{!}{
    \begin{tabular}{c | c c c c c c c c c}
    \toprule
\multirow{2}{*}{Protocol} & \multicolumn{7}{c}{Number of Generated Samples $n$} \\
 & $2^4$ & $2^6$ & $2^8$ & $2^{10}$ & $2^{12}$ & $2^{14}$ & $2^{16}$ & $2^{18}$ \\
\midrule    
ODO-Laplace & 63,360 & 255,360 & 1,021,440 & 4,085,760 & 16,343,040 & 65,372,160 & 261,488,640 & 1,045,954,560 \\
Ostack-Laplace & 13,140 & 18,090 & 37,050 & 102,000 & 336,360 & 1,213,770 & 4,438,140 & 17,319,240 \\
Ostack-Laplace* & 13,140 & 18,090 & 37,050 & 102,000 & 362,370 & 1,213,770 & 4,728,390 & 18,803,850 \\
DNG-Laplace & \textbf{576} & \textbf{2,304} & \textbf{9,216} & \textbf{36,864} & \textbf{147,456} & \textbf{589,824} & \textbf{2,359,296} & \textbf{9,437,184} \\
Transfrom-Laplace & 43,776 & 177,408 & 718,848 & 2,912,256 & 11,796,480 & 47,800,000 & 193,000,000 & 783,000,000 \\
ODO-Gaussian & 1,474,200 & 2,734,920 & 7,037,280 & 22,508,820 & 80,958,960 & 309,042,000 & 1,215,536,400 & 4,852,593,900 \\
Ostack-Gaussian & 74,610 & 98,460 & 179,640 & 420,480 & 1,314,900 & 4,960,890 & 17,555,580 & 68,887,440 \\
DNG-Gaussian & \textbf{576} & \textbf{2,304} & \textbf{9,216} & \textbf{36,864} & \textbf{147,456} & \textbf{589,824} & \textbf{2,359,296} & \textbf{9,437,184} \\
\bottomrule
\end{tabular}
}

\label{tab:lambda-n-bit}
\end{table*}

\subsection{Setup} \label{sec:exp-setup}
Using our benchmark, we conduct comprehensive experiments to compare the efficiency of discrete sampling protocols in various security and privacy settings, which aims to find which protocol to use in the specific setting.  Moreover, we perform a case study using the DP counting query to compare their utility with CDP settings.

We implemented the discrete sampling protocols to sample Laplace and Gaussian noise with MP-SDPZ \cite{keller2020mp} framework using Shamir-BMR \cite{keller2018efficient} with honest majority since it supports three or more computing parties, which enables us to investigate the scalability of sampling protocols. Moreover, the BMR is a binary protocol that efficiently evaluates our benchmark with a large number of bit operations.
Note that MP-SPDZ supports running the same implementation with a variety of underlining protocols. Without changing the code, our benchmark can also be executed with other protocols, such as Yao's protocol and SPDZ.
All the results reported below were run on servers with Intel i7-11700K, Ubuntu 20.04, and 64GB memory. Our settings of range for parameters are $\lambda \in \{64, 128, 256, 512\}$, $\epsilon \in \{0.001, 0.01, 0.1, 1, 10\}$ and $n \in [2^2, 2^{18}]$ and number of computing parties $m \in [2, 8]$, in the setting of $m=2$ we uses Yao's protocol for evaluation. We measure the efficiency of sampling protocols by the number of AND gates, running time and communication, which are the widely used metrics in related works \cite{wei2023securely, champion2019securely, boehler2022secure, bohler2020secure}.


\subsection{Efficiency of  Sampling Protocols}~\label{exp:compare}





In this section, we fix the privacy budget $\epsilon=0.1$, $m=3$ and then set the numbers of generated samples $n$ and security parameters $\lambda$ to compare the efficiency of protocols by the number of AND gates, running time and communication. We also compare the number of required random bits.

\textbf{Efficiency Comparison. } 
The number of AND gates of all the eight sampling protocols in our benchmark are shown in Figure \ref{exp:and}. As for running time and communication in Figure \ref{exp:time} and Figure \ref{exp:com}, we do not include the Trans-Laplace since it has a number of AND gates significantly larger than all the other protocols and it takes days for a single run.
From the experimental results, we have the following observations. (1) The overall trend for all three metrics is the same, i.e., protocols with more AND gates also have longer runtimes and larger communication in real execution. Thus, we mainly use the number of AND gates to present the efficiency in the following several experiments. (2) The AND gates, running time, and communication all increase with the number of samples $n$ and security parameter $\lambda$ for all the sampling protocols, which is intuitive mainly because, as the $n$ and $\lambda$ increase, the number of bits .i.e, $\kappa$, required to represent the noise samples increase for all the protocols. 
Moreover, for the protocol with rejection sampling and Ostack-sampling, the higher $\lambda$ requires larger numbers of Laplace samples $n'$ and CPUSH $u$ are required to reduce the failing rate of generating Laplace samples and biased coins. (3) The ODO-Laplace is the most efficient among the protocols for Laplace noise when $\lambda=128$, while under the large $\lambda$ and $n$, both the Laplace-Ostack and Laplace-Ostack* can be better. Moreover, since Laplace-Ostack* does not require AND get for \proto{RPOP}, it is always better than Laplace-Ostack. These results are consistent with those in \cite{champion2019securely}.  The situation is similar for the Gaussian noise, i.e., the Ostack-based protocol can be superior to the ODO-based one when $\lambda \geq 256$. The reason is that although the Ostack-based method theoretically has lower complexity than the ODO-based one, the CPUSH, and RPOP in Ostack-Sampling (Algorithm \ref{alg:MAKE}) have a number of constant operations, which domain the AND gates especially when $\lambda$ is small. (4) Protocols using DNG with KS testing and uniform transformation are significantly less efficient than protocols using bitwise sampling, especially Transfrom-Laplace, which always has the largest circuit size. That is because Transform-Laplace involves expensive evaluation of logarithmic arithmetic.

\textbf{Random Bits. }
In addition to AND gates, we also evaluate the number of required random bits from the computing parties for the eight protocols in Table \ref{tab:lambda-n-bit}.
We observe that 
Table \ref{tab:lambda-n-bit} shows the number of required random bits under different $\lambda$. We have the following observations. (1) ODO-Gaussian and ODO-Laplace require the largest number of random bits on their corresponding mechanisms since they need $O(l)$ bits to sample a biased coin, while in the Ostack-based protocol, this is reduced to $O(\log{l})$ using the oblivious data structure \cite{champion2019securely}. (2) The random bits used by DNG-based methods DNG-Laplace and DNG-Gaussian are the least and the same since in these two protocols,  the inputs from the computing parties are only partial noise with the same length of $\kappa$. Therefore, DNG-based methods have an advantage over other protocols in the number of random bits required.

\begin{table}[h]
\centering
\caption{Running time (S) of DNG protocols DNG-Laplace, DNG-Gaussian, DNG-Laplace* and DNG-Gaussian*,  the latter two are semi-honest DNG protocols without additional \textbf{Check}. We set $\lambda \in \{128, 256, 512\}$.}
\label{tab:semi-malicious}
\resizebox{0.38\textwidth}{!}{
\begin{tabular}{c | c c c c}
\toprule
\diagbox{$\lambda$}{$\Pi{s}$} & Laplace & Gaussian & Laplace* & Gaussian*  \\
\midrule
$128$ & 1619.07 & 1599.98 & 5.92 & 3.30 \\ 
$256$ & 3239.51 & 1617.30 & 6.20 & 3.32 \\ 
$512$ & 6533.12 & 3219.33 & 6.44 & 3.38 \\ 
\bottomrule
\end{tabular}
}
\end{table}

\textbf{Cost of Malicious Security. }
We also compare the running time of semi-honest DNG protocols with only \texttt{Aggregation} and the Byzantine-resilient DNG protocols, adding \texttt{Check} in Algorithm \ref{alg:test}. We examine both the DNG-Laplace and DNG-Gaussian, setting $\lambda \in \{128, 256, 512\}$. The result is presented in Table \ref{tab:semi-malicious}. We observe that the running time of semi-honest DNG-Laplace* and  DNG-Gaussian* are both significantly lower than that of DNG-Laplace and DNG-Gaussian with additional checks because the semi-honest protocol contains only \texttt{ADD} operations while \texttt{Check} requires iteration over all the samples and for each sample, iterating over the CDF table $\langle F \rangle$. Compared to the results in Figure \ref{fig:exp1_bit_and}, directly aggregating partial noise is markedly more efficient. It takes only seconds to generate 4096 samples even under high security demand. Therefore, in applications requiring only semi-honest security, employing the naive DNG approach that only aggregates partial noise is advisable.

\begin{figure*}[t]
    \centering
    \includegraphics[width=0.99\textwidth]{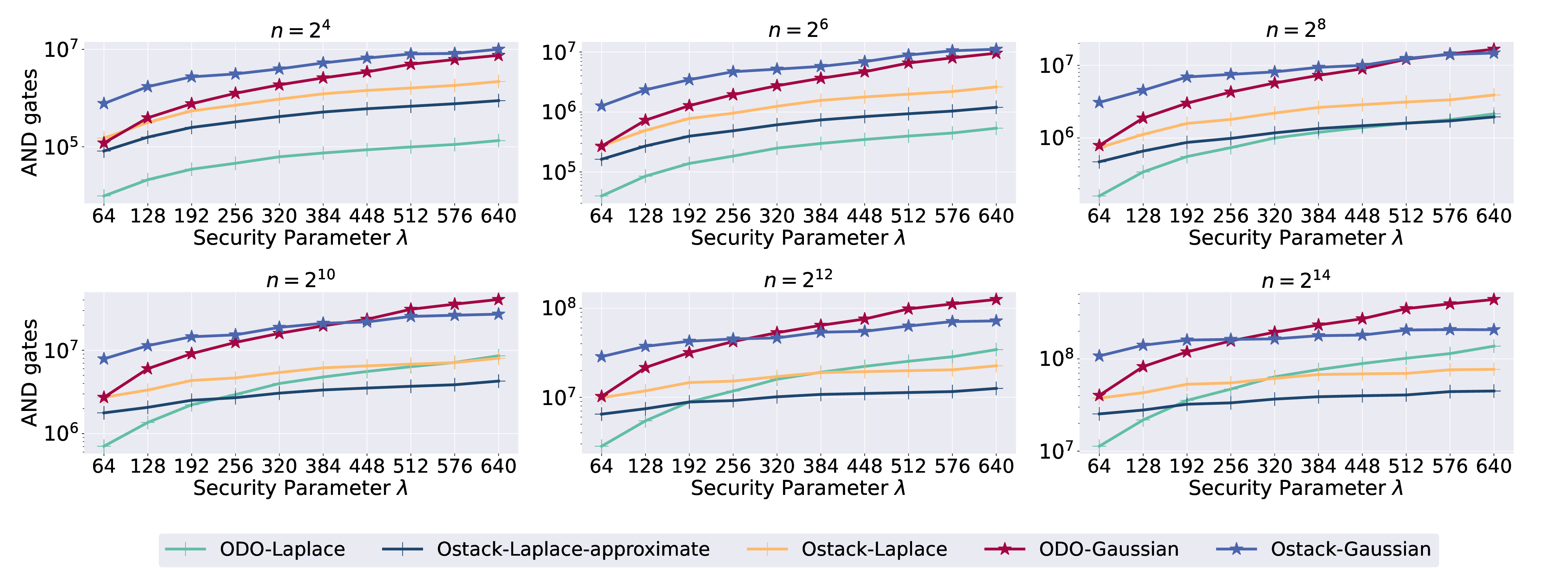}
    \caption{Sampling protocols' trade-off between efficiency and security, measured by security parameter $\lambda$ and AND gates.}
    \label{fig:exp2_and}
\end{figure*}

\begin{figure*}[t]
    \centering
    \includegraphics[width=0.99\textwidth]{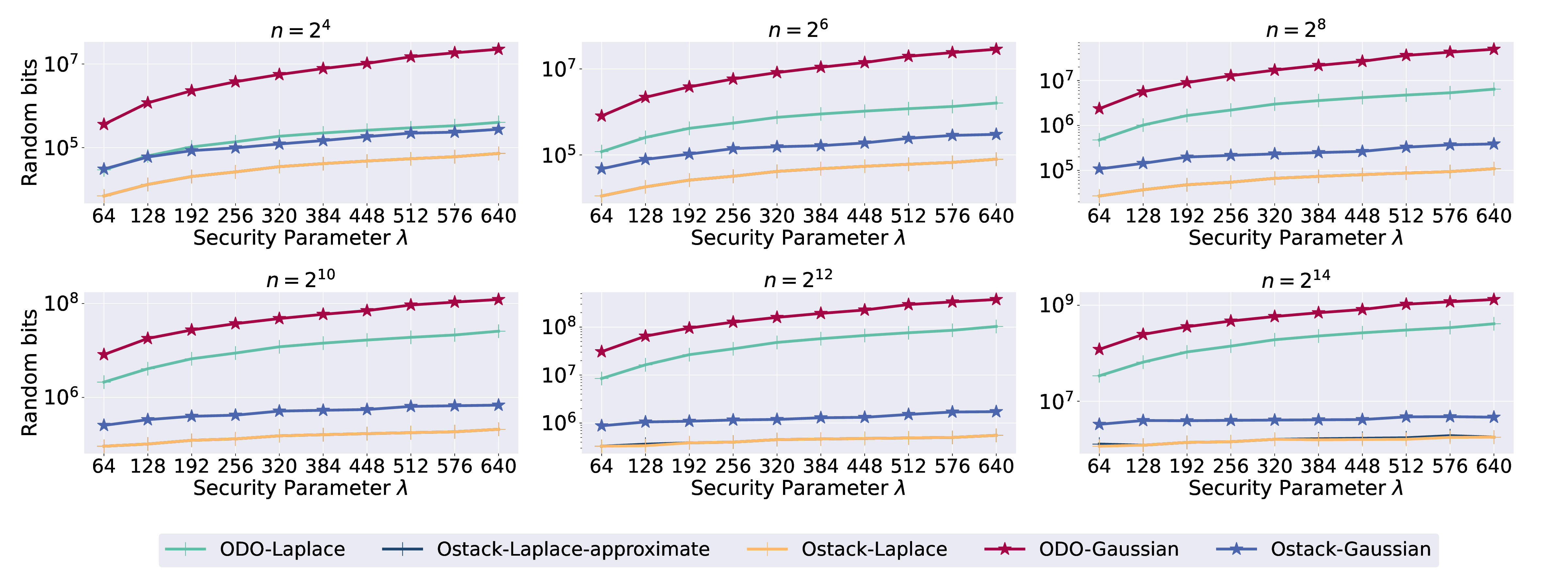}
    \caption{Sampling protocols' trade-off between efficiency and security, measured by security parameter $\lambda$ and random bits.}
    \label{fig:exp2_bits}
\end{figure*}

\subsection{Trade-off between Efficiency and Security of Bitwise Sampling}

Since Bitwise sampling protocols have the least AND gates in all the parameter settings, we conduct further evaluations to give more details by increasing the range of security parameter $\lambda$ to conduct evaluations.
We compare the number of AND gates and random bits of all the bitwise sampling protocols using ODO and Ostack. 

Figure \ref{fig:exp2_and} shows the number of AND gates. For all protocols, the AND gates increase with $\lambda$. 
Specifically, for the Laplace noise, the circuit size of ODO-based methods is smaller than Ostack-based ones when $n < 2^8$. 
And when $n \geq 2^8$, the Ostack-Laplace* has fewer AND gates than ODO-Laplace after $\lambda$ is large enough. 
Moreover, when  $n \geq 2^8$, the Ostack-Laplace is also more efficient than OOD-Laplace. 
On the other hand, the sampling protocols for Gaussian noise also benefit from using Ostack, i.e., the Ostack-Gaussian can have a smaller number of AND gates when $n\geq 2^8$.
We also can clearly see the cross-over points of ODO-based and Ostack-based methods. For the Laplace noise, when $\lambda$ is very small, i.e., 64, the ODO-Laplace always has the fewest AND gates. Furthermore, when $n$ is larger than $n^{10}$, the cross-over point for ODO-Laplace and Ostack-Laplace lays in $256$ and that for ODO-Laplace and Ostack-Laplace* lays in a smaller number, i.e., $192$. As for Gaussian noise, when $n$ is small, the ODO-Gaussian has fewer AND gates than Ostack-Gaussian for all $\lambda$, and when $n$ is large enough, say, larger than $2^{10}$, the cross-over point of ODO-Gaussian and Ostack-Gaussian is also fixed at $\lambda = 256$. It demonstrates that when $\lambda$ and $n$ are small, it is more efficient to use OOD-based sampling methods, while in the situation with high demand for security and a number of noise samples, the Ostack-based methods are more suitable. 
Figure \ref{fig:exp2_bits} presents the random bits input by computing parties. In line with our expectations, the number of random bits required by the ODO-based method is significantly larger than that of the Ostack-based method in all the settings of $\lambda$ and $n$ for both the Laplace and the Gaussian noise.


\begin{figure*}[!t]
    \centering
    \includegraphics[width=1.05\textwidth]{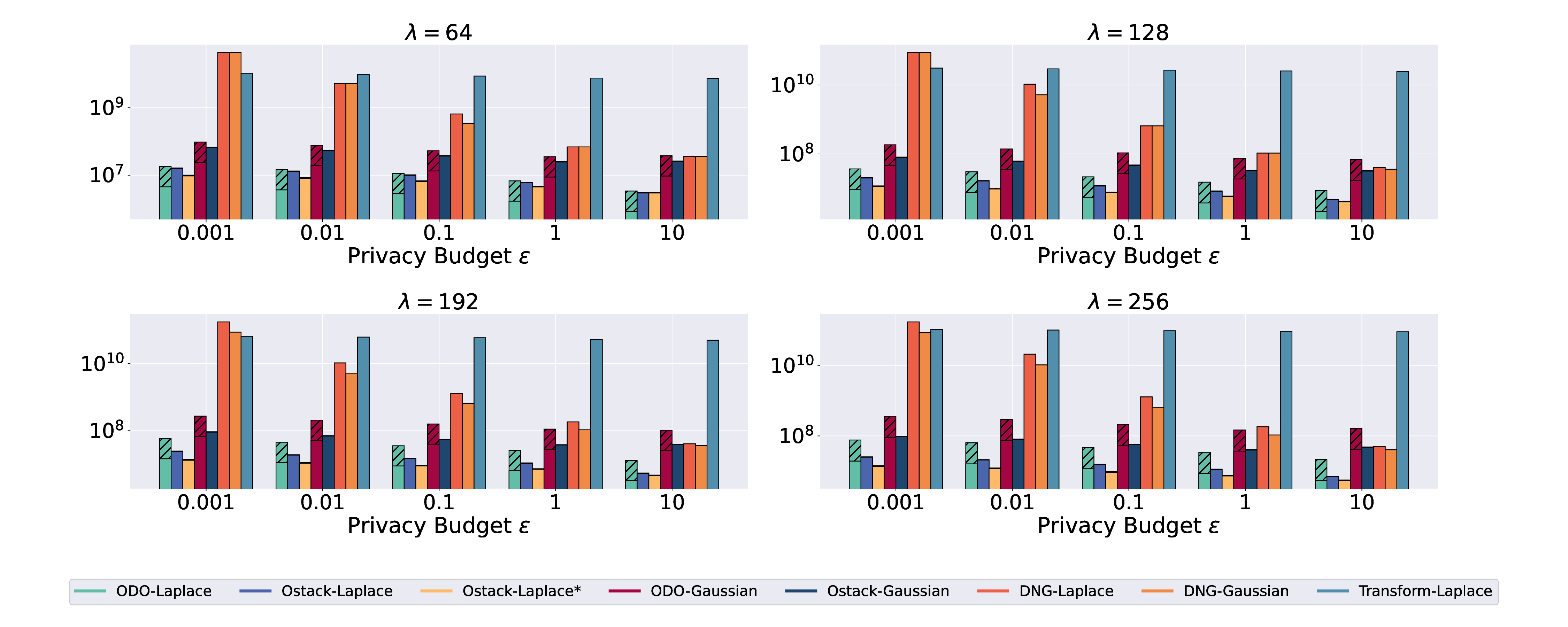}
    \caption{Comparison of sampling protocols’ number of AND gates + random bits under different $\epsilon$ and $\lambda$.}
    \label{fig:AND-eps}
\end{figure*}

\subsection{Efficiency under Various Privacy Demands}

Recall that at larger variances, in order to get the desired statistical distance $\delta_t$, we need to set a larger $N$ in the truncated distribution, which means that the length $\kappa$ of binary expansion used to represent the variable needs to be larger. Therefore, the privacy budget $\epsilon$ can affect all of the eight sampling protocols.
This section explores the impact of privacy budget $\epsilon$ on sampling efficiency.

Figure \ref{fig:AND-eps} presents the number of AND gates in eight sampling protocols with $\epsilon \in \{0.001, 0.01, 0.1, 1, 10\}$ and $\lambda \in \{64, 128, 192, 256\}$.  
We have the following observations. 1) Circuit size increases for all of the eight protocols as $\epsilon$ decreases, which is typical for all the security parameters $\lambda$. 2) The circuit size of Transform-Laplace has the smallest change with $\epsilon$ and stays at a significantly high level. This is because the decimal part domains the length of the fixed-point number used for logarithm arithmetic, and changing the integer part has little impact on the number of AND gates. 3) Moreover, the circuit size of DNG-based protocols has the largest change with $\epsilon$ since the KS-test requires scanning a table of length $N$ to check each sample, resulting in a complexity of $O(N)$.

\subsection{Varying the Number of Computing Parties} \label{sec:num-party}

\begin{figure}[t]
    \centering
    \includegraphics[width=0.45\textwidth]{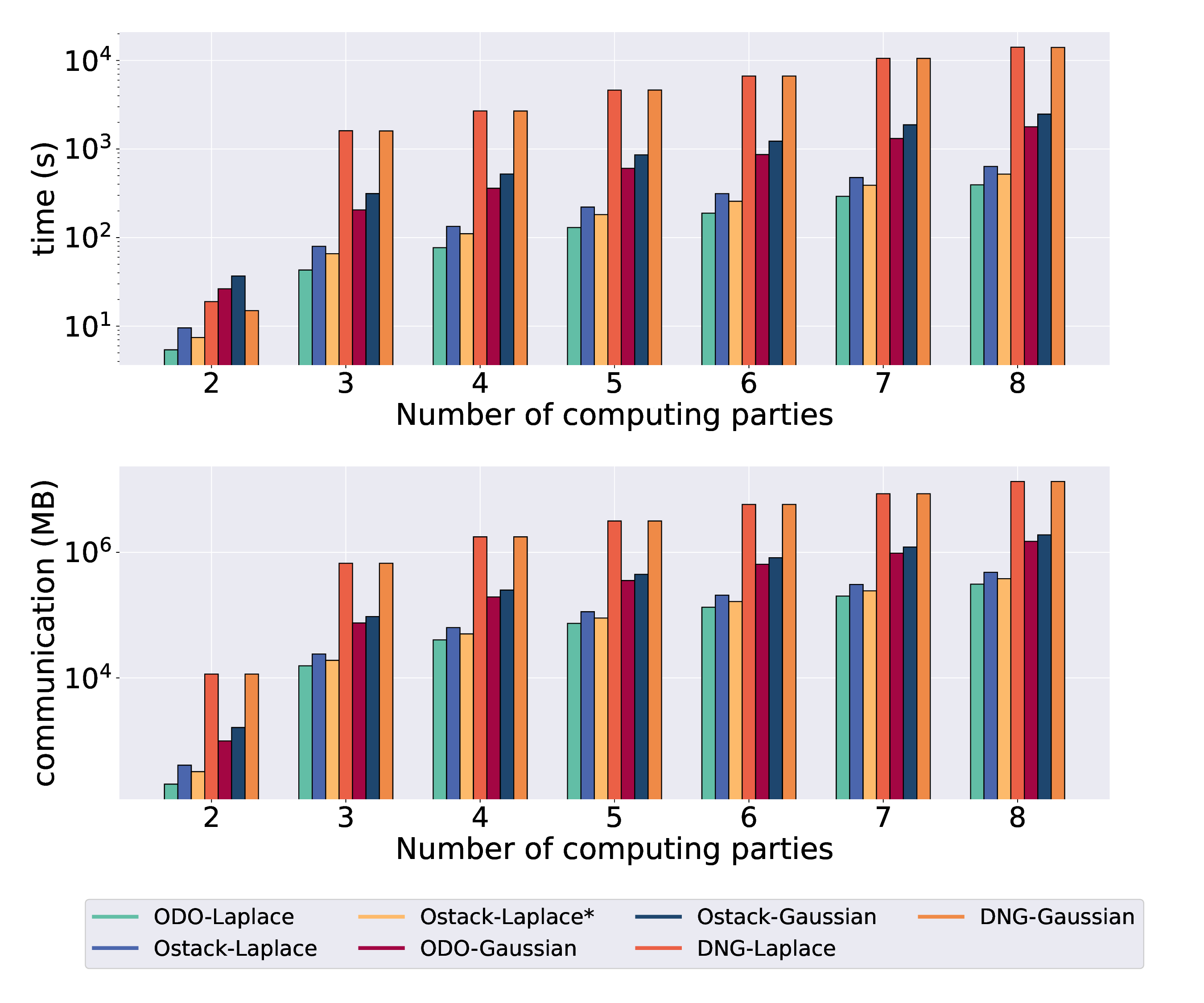}
    \caption{
    The time and communication of protocols under different numbers of parties, where Yao's protocol is used for a two-party setting, and Shamir-BMR is used for others.}
    \label{fig:exp_run_party}
\end{figure}

In this section, we set different numbers of computing parties.
We measure the running time and the communication of seven protocols except Transform-Laplace.
The results are shown in Figure \ref{fig:exp_run_party}. 
We use Yao's protocol to run our code in the two-party setting, and for other $m$, the Shamir-BMR is used.
We observe that both running time and communication increase linearly with the increase of computing party number $m$. 
This demonstrates that the running time of secure sampling protocols is still acceptable in a practical setting with multiple computing parties.
Moreover, Yao's protocol has significantly lower communication and running time since all the protocols can be finished within 100 seconds.

{
\subsection{Utility of Distributed DP: Case Study}

\begin{table*}[h]
\centering
\caption{The MSE of frequency on Kosarak with protocols $\Pi_s$ under different privacy budgets $\epsilon$ and security parameter $\lambda=128$.}
\label{table:eps-uti}
\resizebox{0.86\textwidth}{!}{
\begin{tabular}{c | c c | c c c c c c c | c c }  
\toprule
\diagbox{$\epsilon$}{$\Pi{s}$} & \makecell{CDP\\Laplace} & \makecell{CDP\\Gaussian} & \makecell{ODO\\Laplace} & \makecell{Ostack\\Laplace} & \makecell{Ostack\\Laplace*} & \makecell{DNG\\Laplace} & \makecell{ODO\\Gaussian} & \makecell{Ostack\\Gaussian} & \makecell{DNG\\Gaussian} & \makecell{LDP\\OLH} & \makecell{Shuffle\\SOLH} \\
\midrule
0.1 & 200.03 & 2408.16 & 203.49 & 204.06 & 271.65 & 203.15 & 2380.93 & 2380.93 & 2385.65 & $1.45 \times 10^{13}$ & $6.51 \times 10^{8}$ \\
0.2 & 48.53 & 584.16 & 49.64 & 51.08 & 66.36 & 49.72 & 576.65 & 576.65 & 573.34 & $1.30 \times 10^{13}$ & $4.46 \times 10^{7}$ \\
0.3 & 21.75 & 266.42 & 21.26 & 22.88 & 61.43 & 21.90 & 272.52 & 272.52 & 279.68 & $1.16 \times 10^{13}$ & $1.24 \times 10^{7}$ \\
0.4 & 12.36 & 147.62 & 11.77 & 12.35 & 17.47 & 12.24 & 146.48 & 146.48 & 151.42 & $1.04 \times 10^{13}$ & $7.12 \times 10^{6}$ \\
0.5 & 7.91 & 95.16 & 7.31 & 7.85 & 16.13 & 8.16 & 94.89 & 94.89 & 95.89 & $9.16 \times 10^{12}$ & $5.72 \times 10^{6}$ \\
\bottomrule
\end{tabular}
}
\end{table*}

\begin{table*}[h]
\centering
\caption{The MSE of frequency on Kosarak with protocols $\Pi_s$ under privacy budget $\epsilon=0.1$ and different security parameter $\lambda$.}
\label{table:lambda-uti}
\resizebox{0.75\textwidth}{!}{
\begin{tabular}{c | c | c c c c | c | c c c }
\toprule
\diagbox{$\lambda$}{$\Pi{s}$} & \makecell{CDP\\Laplace} & \makecell{ODO\\Laplace} & \makecell{Ostack\\Laplace} & \makecell{Ostack\\Laplace*}  & \makecell{DNG\\Laplace} & \makecell{CDP\\Gaussian} & \makecell{ODO\\Gaussian} & \makecell{Ostack\\Gaussian} & \makecell{DNG\\Gaussian} \\
\midrule
2 &  \multirow{5}{*}{200.03} & 195.24 & 178.07 & 198.85  & 204.37 & \multirow{5}{*}{2408.16} & 2355.68 & 2355.68 & 2323.40 \\
4 &  & 191.42 & 192.88 & 198.85 & 205.96 & & 2323.81 & 2323.81 & 2350.15 \\
8 &  & 204.35 & 198.85 & 265.94 & 194.98 &  & 2332.28 & 2332.28 & 2350.15 \\
16 &  & 203.06 & 190.69 & 266.95 & 217.22 &  & 2312.47 & 2312.47 & 2286.67 \\
32 &  & 189.71 & 193.03 & 260.05 & 196.97 &  & 2369.28 & 2369.28 & 2401.93 \\
\bottomrule
\end{tabular}
}
\end{table*}


In this section, we validate the utilities of sampling protocols in our benchmark and compare them with center DP, local DP, and shuffle DP under different $\lambda$ and $\epsilon$.
\subsubsection{Setup}
We query the differentially private frequency on the Kosarak dataset\footnote{Kosarak. \href{http://fimi.ua.ac.be/data}{http://fimi.ua.ac.be/data}}
. Kosarak is the click record of the Hungarian news website, which contains about 8 million clicks for $n =41,270$ different web pages. Specifically, we estimate each web page's clicked times. Then, the noise generated in the \textit{central model} (continuous Laplace and Gaussian) and \textit{distributed model} (discrete Laplace and Gaussian) were added to numbers' frequency $f(D)$. We also use OLH ~\cite{wang2017locally} and its shuffle DP version, SOLH~\cite{wang2020improving}, to construct responses on each number (representing single users) to satisfy LDP and shuffle DP. We use MSE to measure and compare the utility of noisy frequency. The Mean Squared Errors is defined as: MSE = $\frac{1}{n} \sum_{i\in[n]} [f(D)-y]^2$. We also show the results in Mean Absolute Error (MAE) and Relative Error (RE) in Appendix~\ref{app:addeval}.

\subsubsection{Comparison Result.}
Table \ref{table:eps-uti} shows the MSE in $\epsilon \in [0.1, 0.5]$ of eight sampling protocols as well as two baseline methods, CDP-Laplace and CDP-Gaussian sampling continuous noise on one server. The results on LDP and shuffle DP are also reported. 
First, we can see that in all $\epsilon$, LDP and shuffle DP have significantly higher MSEs than those of DDP, although they all do not assume a trusted server. Such a large utility gap is due to the random report of the 8 million click records acting as users.
Third, the utilities of DDP protocols are close to those of CDP protocols for both Laplace and Gaussian mechanisms, which matches the expectation for the utility of secure sampling protocols.
Second, the utility (MSE) of continuous CDP-Laplace and Laplace in the DDP model is similar for all $\epsilon$. The only exception is that the Ostack-Laplace* has a larger MSE than other methods because it uses an approximated $\epsilon$, which is the largest number in the form of $2^{-i}\ln{2}, (i\in \mathbb{N})$ smaller than the required one. 
We also fix $\epsilon=0.1$ and set $\lambda=\{1, 2, 4, 8, 16\}$ to measure only the MSE for DDP and CDP protocols in Table \ref{table:lambda-uti}. We observe that the utilities of DDP Laplace protocols under different $\lambda$ are similar. By comparing the experimental results of CDP-Laplace, CDP-Gaussian, and the secure sampling protocols, we can see that the statistical distance caused by MPC in Section~\ref{sec:align} does not introduce additional errors to the answers of the counting query (even when $\lambda$ is very small).

\subsubsection{Utility under Colluded Adversaries and Zero Noise Attack}\label{exp:collude}
As discussed in Section~\ref{sec:semi-dng}, when the parties providing partial noise in DNG collude with each other, by subtracting their noise samples, they may reduce the noise in the final output, which breaks the claimed DP protection of DDP protocols. The impact is identical to the zero noise attacks in Section~\ref{sec:semi-malicious}, where these parties always provide $0$ as their partial noise. To this end, we must ask each party to provide noise with a larger variance.

Assuming different ratios of collusion $\alpha \in [0\%, 90\%]$ ($\alpha = 0\%$ also represents the utility in CDP Gaussian), we use Equation~(\ref{eq:new_variance}) to re-calculate the variance of partial Gaussian noise in DNG-Gaussian. The generation is repeated on $\epsilon \in [0.2, 1.0]$. We add the generated noise to frequencies on Kosarak and show the MSE in Figure~\ref{fig:colluded}. We observe that the impact of colluded attacks is large. Under $90\%$ collusion, the MSE is nearly $100\times$ compared to the non-collusion one (also the utility of CDP Gaussian). Even under the honest majority setting (1/3 collusion), the MSE also increases 5 times. Therefore, for DNG, additional post-processing must be applied to the results (for example, securely shuffling~\cite{chase2020secret} the noisy results before revealing them).

\begin{figure}[t]
    \centering
    \includegraphics[width=0.43\textwidth]{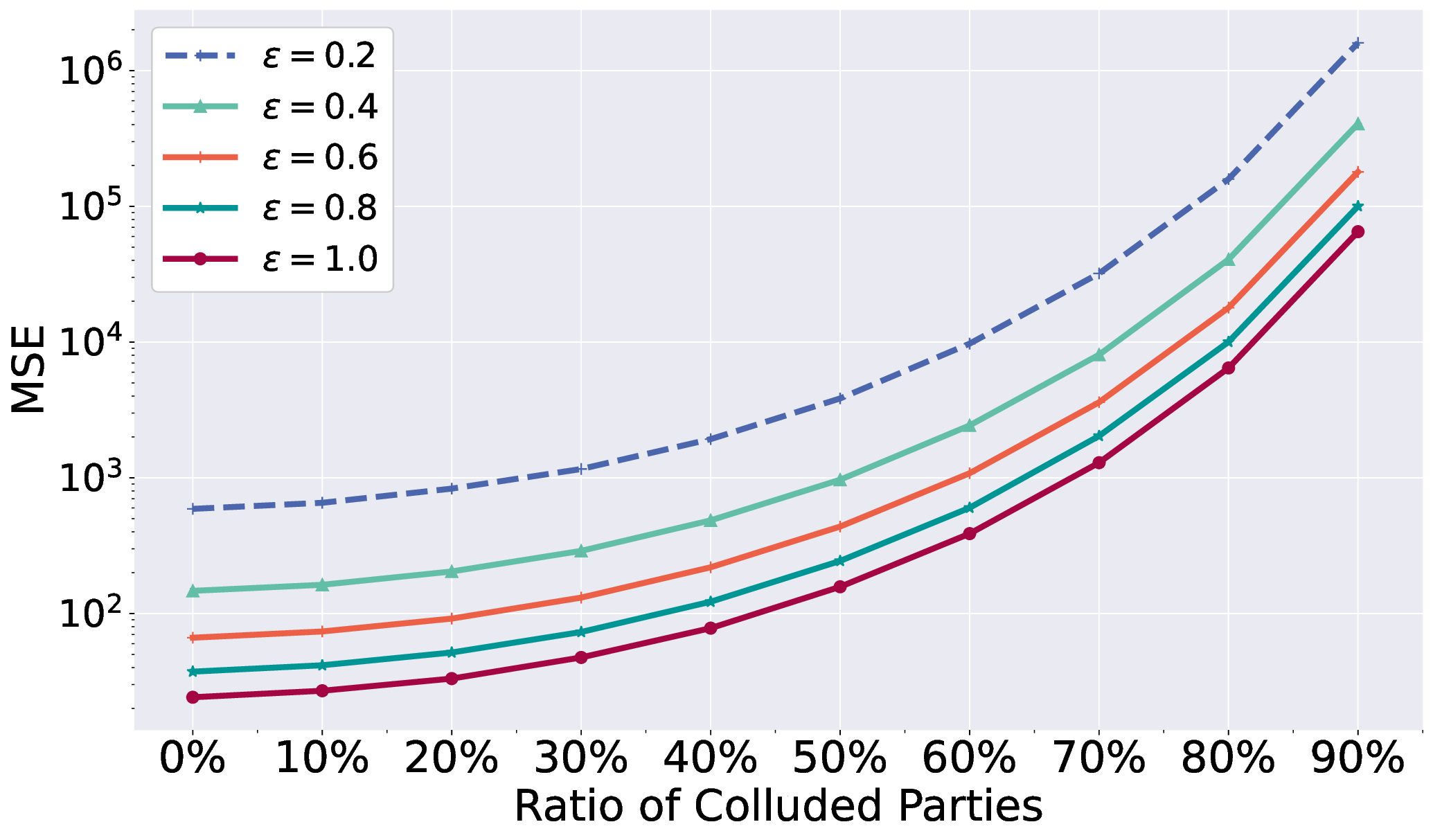}
    \caption{{The MSE of frequency on Kosarak using noise from DNG-Gaussian, assuming ratios of collusion $\alpha \in [0\%, 90\%]$ and privacy budgets $\epsilon \in [0.2, 1.0]$.}}
    \label{fig:colluded}
\end{figure}

}






\section{Takeaways} \label{sec:takeaway} 
This section summarizes the key findings and guidelines considering different factors from our evaluations. 

\mypara{Threat Model.} We first discuss the trade-off on threat models.
\begin{itemize}[leftmargin=*]
    \item \textbf{Semi-honest vs Malicious. } To realize malicious security, the bitwise sampling-based methods are the most effective, while for the semi-honest setting, DNG-based protocols using only \texttt{Aggregation} are the best. That is because the semi-honest DNG-Gaussian involves only \proto{ADD} operations to sum the partial noises. As for the semi-honest DNG-Laplace, it requires additional operations to convert Geometric samples to discrete Laplace samples \cite{wei2023securely}. Thus, the number of AND gates for each sample in these two distributions is $O(1)$. For malicious settings,  the minimum number of AND gates can be found in Ostack-based bitwise sampling for Laplace distribution, which is $O(\log^2(\lambda + \log n))$ \cite{champion2019securely}.
    {
    \item \textbf{Number of Colluded Parties. }
    The fraction of corrupt parties affects the utility of DNG protocols for the Gaussian mechanism since each party should sample Gaussian with additional variances $\hat{\sigma}^2=\sigma^2/(1-\alpha)$, where $\sigma^2$ is the original variance of partial noise and $\alpha$ is the proportion of colluded parties (discussed in Section~\ref{sec:semi-dng}). Thus, the final Gaussian noise also has a variance of $\frac{1}{1-\alpha}$ times compared to the situation without colluded parties.  Note that the bitwise sampling protocols can maintain the same utility because the XOR results of input bits are not revealed.

    \item \textbf{Data Poisoning Attack. } The malicious behavior also exists in the input parties. They can change the local data to skew the final aggregation results in MPC before sampling DDP noise, which is out of the scope of sampling protocols. The data poisoning attack can not be completely addressed. Instead, a promising way is leveraging zero-knowledge proof depending on the specific use case, which ensures the validity of inputs~\cite{bell2023acorn, bell2020secure, roy2022eiffel, boehler2022secure, bohler2020secure}, thus limiting the malicious parties' manipulation of the results. 
    }
\end{itemize}

\mypara{Deployment Model.} We then discuss the deployment models.
\begin{itemize}[leftmargin=*]
    \item \textbf{Statistical Security Parameter $\lambda$. } Now we consider $n > 2^{10}$. To generate Laplace samples with $\lambda < 256 $, we suggest using ODO-Laplace. Otherwise, Ostack-Laplace is more efficient. For the user accepting approximation, with $\lambda > 196$, a more efficient version, i.e., Ostack-Laplapce* can be chosen. However, Ostack-Laplapce* essentially uses smaller $\epsilon$, which has worse utility. As for generating Gaussian samples, with $\lambda<256$, the ODO-Gaussian is suggested, and  Ostack-Gaussian for $\lambda \geq 256$. The suggestion above is from results in Figure \ref{fig:exp2_and}. Note that one can use the predicate function in \cite{champion2019securely} to improve the Ostack-based methods. However, the predicate needs to be hard-coded previously and is hard to re-implement when $\epsilon$ changes.
    \item \textbf{Number of Samples $n$. } 
    {
    In general, the computation overhead and number of input bits are linear in the number of required samples. For example, it needs about 0.2 seconds to generate one discrete Gaussian sample in the malicious setting (Ostack-Gaussian). In machine learning, hundreds of thousands of seconds are needed to generate millions of noise samples, which is not practical. Thus, DNG in the semi-honest setting is usually considered~\cite{10179422}. 
    }
    When generating a small number of samples (when $n \leq 2^8$), using ODO sampling is more efficient. However, when implementing mechanisms with a large demand of Laplace or Gaussian samples like the {Noisy Max Mechanism}~\cite{MT07} (see Appendix \ref{sec:other dp primitive}), it is more suitable to use Ostack-sampling.
    
    { \item \textbf{Number of Parties $m$. } From the experimental results, the running time is linear in the number of computing parties $m$. However, in practice, the computing parties are usually servers with high computing ability, and the number is usually small~\cite{boehler2022secure, wei2023securely} (two or three non-colluded servers). As for the input parties, they only need to secret-share their local data and do not participate in the computation. For $n$ input parties splitting shares to $m$ computing parties, the communication complexity is $O(nm)$.}
    
    \item \textbf{Privacy Budget $\epsilon$. }The privacy budget $\epsilon$ can also affect the efficiency of protocols. Specifically, smaller $\epsilon$ means larger binary expansion length $\kappa$ of generated samples is required to achieve the truncated statistical distance $\delta_t$. As a result, the number of AND gates increases (See Figure \ref{fig:AND-eps}). However, the overall ranking of AND gates is consistent across all protocols, so the choice of protocols is usually determined by $\lambda$ and $n$ rather than $\epsilon$.

\end{itemize}

\mypara{Utility. } The utility of the counting query in DDP is close to that of CDP if the mechanism is realized by discrete Laplace and Gaussian noise, even when setting different security parameters $\lambda$ and privacy budget $\epsilon$ (results in Table \ref{table:eps-uti} and Table \ref{table:lambda-uti}).


\section{Conclusion} \label{sec:con}

This paper presents a benchmark to study the efficiency and utility of various secure sampling protocols to realize differential privacy in the \textit{distributed model}. We first review the existing sampling protocols and present a taxonomy for them. Then, we give further analysis to discuss the security of {distributed noise generation} and present the relationship of various parameters in the {bitwise sampling} methods. We conducted experiments and found that the {bitwise bampling}-based methods are most efficient under various settings since they have the least number of AND gates to evaluate in the binary circuits. Moreover, we also estimate the utility of the DDP counting query on real-world datasets. 

\section*{Acknowledgement}
\noindent We thank all anonymous reviewer construction feedback. Yucheng's work was done while he was an undergrad at Sichuan University and remote interning at the University of Virginia. The work was partially supported by CNS-2220433.



\bibliographystyle{ACM-Reference-Format}
\bibliography{bibliografia.bib}



\begin{center}
    \Large \bf Appendix
\end{center}

\setcounter{section}{0}
\setcounter{equation}{0}
\renewcommand\thesection{\Alph{section}}

\section{Other DP Primitives}
\label{sec:other dp primitive}
\mypara{Exponential Mechanism.}
The exponential mechanism (EM)~\cite{MT07} samples from the set of all possible answers according to an exponential distribution, with answers that are ``more accurate'' being sampled with higher probability.  
This approach requires the specification of a quality function $q$ that takes as input the data $D$, a possible output $o$, and outputs a real-numbered quality score. 
The global sensitivity of the quality functions $\Delta_q$ is defined as:
\begin{align*}
    \Delta_q = \max_{o} \max_{D\simeq D'} |q(D,o) - q(D',o)|
\end{align*} 
The following method ${\rm M}$ satisfies $\epsilon$-differential privacy:
\begin{align}
 \Pr{{\rm M}_q(D)=o} =\frac{ \exp{\left(\frac{\epsilon }{2\,\Delta_q}q(D,o)\right)}}{\sum_o' \exp{\left(\frac{\epsilon }{2\,\Delta_q}q(D,o')\right)}} 
 \label{eq:exp}
\end{align}

As shown in EM~\cite{MT07}, if the quality function is monotonic, i.e., it satisfies the condition that when the input dataset is changed from $D$ to $D'$, the quality scores of all outcomes change in the same direction, i.e., for any neighboring $D$ and $D'$
\begin{align}
\left(\exists_o\, q(D,o) \!<\! q(D',o)\right) \!\implies\! \left(\forall_{o'}\,q(D,o') \!\leq \! q(D',o')\right)\label{eq:monotonicity}    
\end{align}

Then one can remove the factor of $1/2$ in the exponent of \autoref{eq:exp} and return $i$ with probability proportional to $\exp\!{\left(\frac{\epsilon }{\Delta_q}q(D,o)\right)}$.  This improves the accuracy of the result.

\mypara{Noisy Max Mechanism.}
The Noisy Max mechanism (NM)~\cite{dwork2014algorithmic} takes a collection of queries, computes a noisy answer to each query, and returns the index of the query with the largest noisy answer.
More specifically, given a list of quality scores $q(D, o_1), q(D, o_2), \ldots$, the mechanism samples Laplace noises $\Lap{\frac{2\Delta_q}{\epsilon}}$ and adds them to the query scores, i.e., 
\begin{align}
    \tilde{q}(D, o_i) = q(D, o_i) + \Lap{\frac{2\Delta_q}{\epsilon}},\nonumber
\end{align} 
and returns the index $\max_i\tilde{q}(D, o_i)$.
Dwork and Roth prove this satisfies $\epsilon$-DP~\cite{dwork2014algorithmic}.  Moreover, if the queries satisfy the monotonic condition (i.e., Eq.~\eqref{eq:monotonicity}), one can remove the factor of $2$ in the Laplace noise, improving the result's accuracy.  Note that in addition to Laplace noise here, we can also add Gumbel noise (which is equivalent to EM) and exponential noise (only the positive side of Laplace noise).
Ding et al. \cite{ding2023floating} also propose an extension of Noise Max that can publish top-$k$ and with free gap information in a finite-precision secure implementation taking advantage of integer arithmetic. Champion et al. \cite{champion2019securely} implement the Noisy Max mechanism within the MPC protocol.  


\mypara{Sparse Vector Technique.}
Similar to the Exponential mechanism and Report-Noisy-Max, the Sparse Vector Technique (SVT) is also a method for selection problems.  Unlike the first two, which operate in one shot (i.e., they are given the quality function $q$ and directly output an output $o$), SVT executes sequentially: it evaluates the quality function on each possible $o$ one by one, and after each evaluation (SVT also adds noise to each query result), it outputs a symbol indicating whether the result is high enough or not.  Compared to the Exponential mechanism and Report-Noisy-Max, the sequential style of SVT has both benefits and drawbacks: it allows adaptivity/interactivity, which means one can update the query function during the execution of SVT (after observing each result); on the other hand, because SVT does not have the global view before the end, at each step, it has to make a guess of whether the current noisy result is the highest one, making it less accurate.
Before diving into details, we make a notational change for representation purposes: the quality function $q(\Data, o)$ can be represented as a sequence of queries $q_i$ to $\Data$.  Here, we replace $o$ with $i$, and each $q(\Data, o)$ is changed to $q_i(\Data)$.  The two sets of notations are equivalent and can be used interchangeably.
Recently, Zhu et al. \cite{zhu2020improving} proved that a variant of SVT by adding Gaussian noise can have a better utility than the standard SVT. 

Note that all the mechanisms mentioned above can be reduced to the key step of sampling Laplace or Gaussian noise.

\begin{algorithm}[t]
\caption{ODO-Sampling $(n, l, p)$  \cite{dwork2006our}}
\label{alg:ODO}
{
\begin{algorithmic}[1]
\INPUT Number of required Bernoulli samples $n$, binary expansion of bias $p$ of length $l$
\OUTPUT Shares of samples $\langle x_1 \rangle, ..., \langle x_n \rangle$ from Bernoulli distribution $\mathcal{B}(p)$
\FOR{$i:=1$ to $n$}
\STATE $\langle x_i \rangle = \langle 1 \rangle$
    \FOR{$j:=l$ to $1$}
        \STATE $\langle b \rangle = \proto{URBIT}()$
        \STATE $\langle t \rangle = \proto{XOR}(p_i, \langle b \rangle)$
        \STATE $\langle x_i \rangle = \proto{MUX}(\langle t \rangle, \neg \langle b \rangle, \langle x_i \rangle)$
    \ENDFOR
\ENDFOR
\RETURN $\langle x_1 \rangle, ..., \langle x_n \rangle$
\end{algorithmic}
}
\end{algorithm}

\begin{algorithm}[t]
\caption{Ostack-Sampling $(g, u, l, p)$ \cite{champion2019securely}}
\label{alg:MAKE}
{
\begin{algorithmic}[1]
\INPUT Size of Ostack $g$ to store Bernoulli samples, number of PUSH $u$, binary expansion of bias $p$ of length $l$
\OUTPUT Shares of samples $\langle x_1 \rangle, ... \langle x_g \rangle$ from Bernoulli distribution $\mathcal{B}(p)$
\STATE Find the minimum $r=3\cdot (2 ^ i - 1)$ such that $r \geq l$.
\STATE Initialize an OStack ${\rm rstack}_{r}$ to store $p$.
\STATE Initialize an Ostack ${\rm cstack}_{g}$ to store Bernoulli samples.
\FOR{$i:=1$ to $u$}
\STATE $\langle b \rangle = \proto{URBIT}()$
\STATE $\langle p' \rangle = \proto{RPOP}({\rm rstack}_{r})$
\STATE $\langle t \rangle = \proto{XOR}(\langle p' \rangle, \langle b \rangle)$
\STATE $\proto{CPUSH}(\langle t \rangle , \neg \langle b \rangle, {\rm cstack}_{g})$
\STATE $\proto{CRESET}(\langle t \rangle, {\rm rstack}_{r})$
\ENDFOR
\STATE $\langle x_1 \rangle, ..., \langle x_g \rangle = \proto{PURGE}({\rm cstack}_{g})$
\RETURN $\langle x_1 \rangle, ..., \langle x_n \rangle$
\end{algorithmic}
}
\end{algorithm}

\begin{table}[t]
    \caption{The definition of some probability distribution. Here $c$ in PDF represents the normalized constant.}
    \label{dist}
    \centering
    \resizebox{0.49\textwidth}{!}{
    \begin{tabular}{cc}
    \toprule
         Distribution & PDF \\
    \midrule
          Bernoulli  $\mathcal{B}(p)$ & $f_\mathcal{B}(0)=1-p$,  $f_\mathcal{B}(1)=p$\\
          Uniform   $\mathsf{U}(a,b)$ & $f_\mathsf{U}(x)=\frac{1}{b-a}$ for $x\in[a,b]$\\
          Discrete Uniform   $\mathsf{U}_\mathbb{Z}(a,b)$ & $f_{\mathsf{U}_\mathbb{Z}}(x)=\frac{1}{b-a}$ for $x\in[a,b]\cap \mathbb{Z}$\\   
          Laplace   $\mathsf{Lap}(t)$ & $f_\mathsf{Lap}(x)=c\cdot e^{-|x|/t}$ for $x\in\mathbb{R}$\\
          Gaussian   $\Gau{\sigma^2}$ & $f_\mathsf{N}(x)=c\cdot e^{-x^2/(2\sigma^2)}$ for $x\in\mathbb{R}$\\
    \midrule
          Discrete  Laplace   $\dLap{t}$ & $f_{\mathsf{Lap}_\mathbb{Z}}(x)=c\cdot e^{-|x|/t}$ for $x\in\mathbb{Z}$, \\
          Discrete  Gaussian  $\dGau{\sigma^2}$ & $f_{\mathsf{N}_{\mathbb{Z}}}(x)=c\cdot e^{-x^2/(2\sigma^2)}$ for $x\in\mathbb{Z}$ \\    
          Negative Binomial  $\mathsf{NB}(r, p)$ & $f_{\mathsf{NB}}(x) = \binom{r+x-1}{r-1}p^x(1-p)^r$ for $x\in\mathbb{N}$\\  
          Gamma   $\mathsf{GM}(k, b)$ & $f_{\mathsf{GM}}(x) = \frac{1}{\Gamma(k)b^k}x^{k-1}e^{-\frac{x}{b}}$ for $x\in\mathbb{R}$ \\  
    \bottomrule
    \end{tabular}
    }
\end{table}

\begin{table}[t]
\centering
\caption{Some MPC subprotocols mentioned in this paper. }
\label{tab:subprotocol}
\begin{tabular}{cc}
\toprule
MPC subprotocol & Output / Functionality \\
\midrule
$\proto{EQ}(\langle a \rangle ,\langle b \rangle)$  & If $a=b$, $\langle 1 \rangle$, else $\langle 0 \rangle$ \\
$\proto{LE}(\langle a \rangle ,\langle b \rangle)$ & If $a\leq b$, $\langle 1 \rangle$, else $\langle 0 \rangle$ \\
$\proto{ADD}(\langle a \rangle ,\langle b \rangle)$  & $\langle a + b \rangle$ \\
$\proto{SUB}(\langle a \rangle ,\langle b \rangle)$  & $\langle a - b \rangle$ \\
$\proto{ABS}(\langle a \rangle)$  & $\langle |a| \rangle$ \\
$\proto{XOR}(\langle a \rangle ,\langle b \rangle)$ & $\langle a \oplus b \rangle$ \\
$\proto{MUX}(\langle a \rangle ,\langle b \rangle, \langle c \rangle)$ & If $a=1$, $\langle b \rangle$, else $\langle c \rangle$ \\
$\proto{UBIT}()$ \cite{wei2023securely} & A uniform random bit $\langle b \rangle$\\
$\proto{SORT}(\langle A \rangle) $ \cite{keller2020mp} & Securely sort list $\langle A \rangle$ \\
\bottomrule
\end{tabular}
\end{table}

\begin{table*}
    \centering
    \caption{Summary of partial noise in local sampling and the arithmetics in aggregation.}
    \label{tab:summary-dng}
    \begin{tabular}{c|ccc}
        \toprule
        Targeted Distribution & Partial Distribution & Local Sampling & Aggregation \\
        \midrule
        \multirow{3}{*}{Continuous Laplace} & Gamma & $Y_{i,1}, Y_{i,2} \sim {\rm Gamma} (\frac{1}{m}, \frac{\Delta}{\epsilon})$ & $\langle L \rangle = \sum_{i=1}^m (\langle Y_{i,1} \rangle - \langle Y_{i,2} \rangle)$ \\
        & Gaussian & $N_{i,1}, N_{i,2}, N_{i,3}, N_{i,4} \sim  \Gau{\frac{\Delta}{2m \epsilon}}$ & 

        $\langle N_j \rangle = \sum_{i=1}^m \langle N_{i,j} \rangle$, $\langle L \rangle = \sum_j^4 (-1)^j \langle N_j \rangle$ \\

        & Laplace and Beta &  $B \sim \mathcal{B}(1, m-1)$, $L_i \sim \Lap{\frac{\Delta}{\epsilon}}$ & $\langle L \rangle = \langle \sqrt{B} \rangle \cdot \sum_{i=1}^m \langle L_i \rangle$ \\
        \midrule
        Discrete Laplace & Negative Binomial & $Y_{i} \sim  \mathsf{NB} (1/m, e^{-\frac{\Delta}{\epsilon}}) $ & $\langle L \rangle = \sum_{i=1}^m \langle Y_i \rangle $\\
        \midrule
        Gaussian & Gaussian & $Y_i \sim \Gau{\frac{2\ln{1.25/\delta} \Delta^2}{m\epsilon^2}} $ & $ \langle G \rangle = \sum_{i=1}^m \langle Y_i \rangle $\\
        \bottomrule
    \end{tabular}
    \label{tab:my_label}
\end{table*}

\begin{figure*}[!t]
    \centering

    \begin{subfigure}[b]{0.99\textwidth}
    \includegraphics[width=\textwidth]{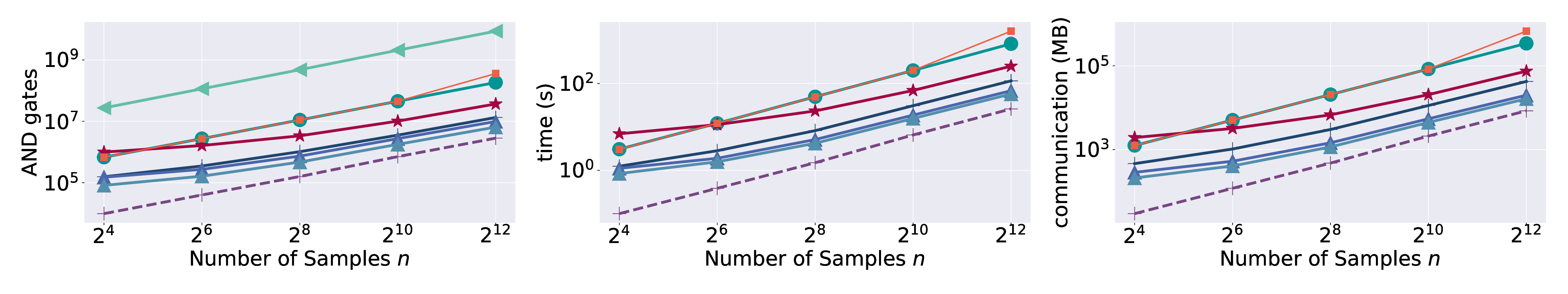}
    \end{subfigure}

    \begin{subfigure}[b]{0.99\textwidth}
        \includegraphics[width=\textwidth]{figures/exp1/and+bit/legend.eps}
    \end{subfigure}

    \caption{Number of AND gates, running time, and communication in Shamir-BMR in $\lambda=64$ and different $n$.}
    \label{fig:exp1_bit_and_64}
\end{figure*}

\begin{table*}[t]
\centering
\caption{The MAE and RE of counting using various sampling protocol $\Pi_s$ with DP budgets $\epsilon$ and security parameter $\lambda=128$.}
\label{table:eps-uti-app}
\resizebox{\textwidth}{!}{
\begin{tabular}{c | c | c c c c c | c c c c }
\toprule
Metric & \diagbox{$\epsilon$}{$\Pi{s}$} & CDP-Laplace & ODO-Laplace & Ostack-Laplace & Ostack-Laplace*  & DNG-Laplace & CDP-Gaussian & ODO-Gaussian & Ostack-Gaussian & DNG-Gaussian \\
\midrule
\multirow{5}{*}{MAE} & 0.1 & 10.08 & 10.05 & 10.05 & 11.49 & 9.94 & 38.74 & 38.95 & 39.48 & 38.92 \\
& 0.2 & 4.97 & 4.92 & 4.95 & 5.62 & 4.97 & 19.03 & 19.13 & 19.13 & 19.22 \\
& 0.3  & 3.42 & 3.24 & 3.29 & 5.21 & 3.28 & 12.75 & 13.12 & 13.12 & 13.32 \\
& 0.4 & 2.44 & 2.35 & 2.43 & 2.83 & 2.41 & 9.60 & 9.64 & 9.64 & 9.80 \\
& 0.5 & 2.03 & 1.85 & 1.91 & 2.64 & 1.97 & 7.87 & 7.77 & 7.77 & 7.79 \\
\midrule
\multirow{5}{*}{RE} & 0.1 &  334.38 & 345.51 & 325.48 & 395.04 & 325.63 & 1276.84 & 1320.62 & 1298.97 & 1301.34 \\
& 0.2  & 165.45 & 165.66 & 163.96 & 187.05 & 164.60 & 625.88 & 632.52 & 632.52 & 637.33 \\
& 0.3  & 115.40 & 106.68 & 108.97 & 173.11 & 107.38 & 432.43 & 440.82 & 440.82 & 446.43 \\
& 0.4 & 82.34 & 77.94 & 80.91 & 93.45 & 82.36 & 321.25 & 319.64 & 319.64 & 321.06 \\
& 0.5 & 67.44 & 59.38 & 64.45 & 87.77 & 61.81 & 263.29 & 257.97 & 257.97 & 255.21 \\
\bottomrule
\end{tabular}
}
\end{table*}

\section{Protocols Sampling Biased Coins} \label{app:sampling}

In this section, we give the subprotocols ODO-sampling \cite{dwork2006our} and Ostack-sampling \cite{champion2019securely} for generating biased coins in bitwise sampling protocols.

\section{Notations and Subprotocols}
\label{app:notations}

We use $\mathbb{Z},\mathbb{N},\mathbb{R}$ to denote the set of integers, non-negative integers, and real numbers, respectively. For function $f$, we denote $f:\mathbb{F}\rightarrow\mathbb{O}$ if $f$ has input domain $\mathbb{F}$ and output domain $\mathbb{O}$. For distribution $\mathcal{X}$, we denote $f_\mathcal{X}(\cdot)$ as the probability density function (PDF) of $\mathcal{X}$, and $F_\mathcal{X}(\cdot)$ as the cumulative distribution function (CDF) of $\mathcal{X}$.

Table \ref{dist} shows the definition of some probability distribution used in this paper, and Table~\ref{tab:subprotocol} describes some subprotocols along with output/functionality.

\section{Summary of DNG} \label{app:dng}

The partial noise in local sampling and the arithmetics in secure aggregation for DNG protocols are shown in Table \ref{tab:summary-dng}.

\section{Additional Evaluations}\label{app:addeval}

This section gives more evaluation results. We show the number of AND gates, running times, and communication of the sampling protocols when $\lambda=64$ in Figure \ref{fig:exp1_bit_and_64}, and the results with larger $\lambda$ have been shown in Figure \ref{fig:exp1_bit_and}.
We also present the MAE and RE of counting using different Sampling protocols in Table~\ref{table:eps-uti-app} as supplements. The Mean Absolute Error (MAE) is defined as $\frac{1}{n} \sum_{i\in[n]} |f(D)-y|$.  The Relative Error (RE) is defined as $\frac{1}{n} \sum_{i\in[n]} | \frac{f(D)-y}{f(D)} | \times 100\% $.

\end{document}